\documentclass{jpp}

\usepackage{graphicx}
\usepackage{natbib}

\usepackage{amsmath}
\usepackage{amssymb}
\usepackage[makeroom]{cancel}
\usepackage{color}

\ifCUPmtlplainloaded \else
  \checkfont{eurm10}
  \iffontfound
    \IfFileExists{upmath.sty}
      {\typeout{^^JFound AMS Euler Roman fonts on the system,
                   using the 'upmath' package.^^J}%
       \usepackage{upmath}}
      {\typeout{^^JFound AMS Euler Roman fonts on the system, but you
                   dont seem to have the}%
       \typeout{'upmath' package installed. JPP.cls can take advantage
                 of these fonts, if you use 'upmath' package.^^J}%
      }
  \else
  \fi
\fi

\ifCUPmtlplainloaded \else
  \checkfont{msam10}
  \iffontfound
    \IfFileExists{amssymb.sty}
      {\typeout{^^JFound AMS Symbol fonts on the system, using the
                'amssymb' package.^^J}%
       \usepackage{amssymb}%
         
       \let\ge=\geqslant  
      }{}
  \fi
\fi

\ifCUPmtlplainloaded \else
  \IfFileExists{amsbsy.sty}
    {\typeout{^^JFound the 'amsbsy' package on the system, using it.^^J}%
     \usepackage{amsbsy}}
    {\providecommand\boldsymbol[1]{\mbox{\boldmath $##1$}}}
\fi

\newcommand{\figref}[1]{figure~\ref{#1}}
\newcommand{\figsref}[1]{figures~\ref{#1}}
\newcommand{\Figref}[1]{Figure~\ref{#1}}

\newcommand{\Secref}[1]{Section~\ref{#1}}
\newcommand{\secref}[1]{section~\ref{#1}}

\newcommand{\secsand}[2]{sections~\ref{#1} and \ref{#2}}

\newcommand{\apref}[1]{appendix~\ref{#1}}

\newcommand{\Eqref}[1]{Equation~(\ref{#1})}
\renewcommand{\eqref}[1]{equation~(\ref{#1})}
\newcommand{\Eqsref}[1]{Equations~(\ref{#1})}
\newcommand{\eqsref}[1]{equations~(\ref{#1})}
\newcommand{\Eqsand}[2]{Equations~(\ref{#1}) and (\ref{#2})}
\newcommand{\eqsand}[2]{equations~(\ref{#1}) and (\ref{#2})}

\newcommand{\exref}[1]{(\ref{#1})}

\newcommand{\bea}{\begin{eqnarray}}
\newcommand{\eea}{\end{eqnarray}}
\newcommand{\beq}{\begin{equation}}
\newcommand{\eeq}{\end{equation}}
\newcommand{\lt}{\left}
\newcommand{\rt}{\right}

\newcommand{\la}{\langle}
\newcommand{\ra}{\rangle}

\newcommand{\mbf}[1]{\mathbf{#1}}

\newcommand{\dd}{\partial}
\newcommand{\vdel}{\boldsymbol{\nabla}}
\newcommand{\vdperp}{\vdel_\perp}

\newcommand{\dpar}{\nabla_\parallel}

\newcommand{\const}{\mathrm{const}}

\newcommand{\vr}{\mbf{r}}
\newcommand{\vrperp}{\vr_\perp}
\newcommand{\rperp}{r_\perp}
\newcommand{\rperpc}{r_{\perp\mathrm{c}}}
\newcommand{\vz}{\hat{\mbf{z}}}

\newcommand{\vk}{\mbf{k}}
\newcommand{\vp}{\mbf{p}}

\newcommand{\vkperp}{\vk_\perp}
\newcommand{\kperp}{k_\perp}
\newcommand{\kxo}{k_{x 0}}
\newcommand{\kyo}{k_{y 0}}
\newcommand{\kperpo}{k_{\perp 0}}
\newcommand{\kperpc}{k_{\perp c}}

\newcommand{\vpperp}{\vp_\perp}
\newcommand{\pperp}{p_\perp}

\newcommand{\kpar}{k_\parallel}
\newcommand{\kparo}{k_{\parallel0}}
\newcommand{\kparc}{k_{\parallel\mathrm{c}}}

\newcommand{\ppar}{p_\parallel}
\newcommand{\qpar}{q_\parallel}
\newcommand{\Lpar}{L_\parallel}
\newcommand{\Lperp}{L_\perp}
\newcommand{\DT}{D_{\mathrm{turb}}}

\newcommand{\tauco}{\tau_{\mathrm{nl}0}}
\newcommand{\tauc}{\tau_{\mathrm{nl}}}
\newcommand{\mc}{m_{\mathrm{c}}}
\newcommand{\scoll}{s_{\mathrm{c}}}

\newcommand{\vv}{\mbf{v}}
\newcommand{\vvperp}{\vv_\perp}
\newcommand{\vperp}{v_\perp}
\newcommand{\vpar}{v_\parallel}
\newcommand{\hvpar}{\hat v_\parallel}
\newcommand{\vth}{v_{{\rm th}}}

\newcommand{\vthe}{v_{{\rm th}e}}
\newcommand{\mfp}{\lambda_\mathrm{mfp}}

\newcommand{\vu}{\mbf{u}}
\newcommand{\vuperp}{\vu_\perp}

\newcommand{\uperp}{u_\perp}
\newcommand{\uperpo}{u_{\perp 0}}

\newcommand{\upar}{u_\parallel}

\newcommand{\dTpar}{\delta T_\parallel}

\newcommand{\vE}{\mbf{E}}
\newcommand{\vB}{\mbf{B}}

\newcommand{\dn}{\delta n}

\newcommand{\tf}{\tilde f}
\newcommand{\tg}{\tilde g}
\renewcommand{\tt}{\tilde t}

\newcommand{\eps}{\varepsilon}
\newcommand{\ephi}{\varphi}
\newcommand{\ephio}{\ephi_0}

\newcommand{\Eph}{E_\ephi}

\newcommand{\FM}{F_{\mathrm{M}}}
\newcommand{\df}{\delta f}
\newcommand{\rmd}{\mathrm{d}}
\newcommand{\sgn}{\mathrm{sgn}}
\renewcommand{\Re}{\mathrm{Re}}
\renewcommand{\Im}{\mathrm{Im}}

\title[Phase mixing vs.\ nonlinear advection in plasma turbulence]{Phase mixing vs.\ nonlinear
advection in drift-kinetic plasma turbulence}

\author[A.\ A.\ Schekochihin et al.]%
{A.~A.~Schekochihin,$^{1,2}$%
\thanks{Email: alex.schekochihin@physics.ox.ac.uk}  
J.~T.~Parker,$^{3,4}$
E.~G.~Highcock,$^{1,4}$
P.~J.~Dellar,$^{3}$
W.~Dorland$^{5,2,1}$
and G.~W.~Hammett$^{6,2,1}$
}

\affiliation{
$^1$Rudolf Peierls Centre for Theoretical Physics, University of Oxford, 1 Keble Road,\\ 
Oxford OX1 3NP, UK
\\[\affilskip]
$^2$Merton College, Merton Street, Oxford OX1 4JD, UK
\\[\affilskip]
$^3$OCIAM, Mathematical Institute, University of Oxford, Andrew~Wiles Building,\\ 
Radcliffe Observatory Quarter, Woodstock Road, Oxford OX2 6GG, UK 
\\[\affilskip]
$^4$Brasenose College, Radcliffe Square, Oxford OX1 4AJ, UK
\\[\affilskip]
$^5$Department of Physics, University of Maryland, College Park, Maryland 20742, USA
\\[\affilskip]
$^6$Plasma Physics Laboratory, Princeton University, P.~O.~Box~451,\\ 
Princeton, New Jersey 08543, USA}

\begin{document}

\maketitle

\begin{abstract}
A scaling theory of long-wavelength electrostatic turbulence in a magnetised, weakly collisional plasma 
(e.g., drift-wave turbulence driven by ion temperature gradients) is proposed, 
with account taken both of the nonlinear advection of the perturbed particle distribution 
by fluctuating $\vE\times\vB$ flows and of its phase mixing, which is caused by the
streaming of the particles along the mean magnetic field and, in a linear problem, would 
lead to Landau damping. It is found that it is possible 
to construct a consistent theory in which very little free energy leaks into high 
velocity moments of the distribution function, rendering the turbulent cascade in 
the energetically relevant part of the wave-number space essentially fluid-like. 
The velocity-space spectra of free energy expressed in terms of Hermite-moment orders 
are  steep power laws and so the free-energy content of the phase space does not diverge 
at infinitesimal collisionality (while it does for a linear problem); collisional
heating due to long-wavelength perturbations vanishes in this limit (also in contrast 
with the linear problem, in which it occurs at the finite rate equal to the 
Landau-damping rate). The ability of the free energy to stay in the low 
velocity moments of the distribution function is facilitated by the  
``anti-phase-mixing'' effect, whose presence in the nonlinear system is 
due to the stochastic version of the plasma echo 
(the advecting velocity couples the phase-mixing and anti-phase-mixing perturbations). 
The partitioning of the wave-number space between the (energetically dominant) region 
where this is the case and the region where linear phase mixing wins its competition 
with nonlinear advection is governed by the ``critical balance'' between linear 
and nonlinear timescales (which for high Hermite moments splits into two 
thresholds, one demarcating the wave-number region where phase mixing predominates, 
the other where plasma echo~does). 
\end{abstract}


\section{Introduction} 

Turbulence is a process whereby energy injected into a system (via some 
mechanism usually associated with the system being out of equilibrium) 
is transferred nonlinearly---and therefore leading to chaotic and multiscale 
states---from the scale(s) at which it is injected to much smaller scales 
at which it is thermalised through microphysical dissipation channels available 
in the system. The system is forced to seek ways of transferring 
energy across a range of scales because the injection and dissipation physics are usually 
unrelated to each other 
and operate at disparate scales. It is the bridging of the gap between these scales 
that brings about turbulent cascades, broad-range power-law spectra, and so on. 
In fluid systems, however varied and multi-physics they are, most turbulence 
theories are basically extensions and generalisations of the ideas 
of \citet{richardson22} and \citet{K41} of a local-in-scale cascade 
maintaining a constant flux of energy away from the injection and towards 
the dissipation scales \citep[e.g.,][]{zakharov92,davidson13}. 
This type of thinking has been tremendously successful in making 
sense of experimental and numerical evidence in both fluids and plasmas.  

In plasmas, however, a straightforward application of such ``fluid'' thinking 
to any physical regime that is not collisionally dominated skirts over the obvious 
complication that the kinetic phase space includes the particle velocities 
as well as their positions, and the (free) energy is generally free to travel 
across this entire 6D space. Its ability---and propensity---to do so is, 
in fact, manifest in what is probably the most important phenomenon that 
makes plasmas conceptually different from fluids---the \citet{landau46} 
damping of electromagnetic perturbations in a collisionless plasma. 
Viewed in energy terms, it involves the transfer of 
free energy from electromagnetic perturbations into perturbations of the particle 
distribution function, which develops ever finer structure in velocity 
space (``phase mixing'') until this transfer (which looks like damping if one 
only tracks the electromagnetic fields) is made irreversible by coarse graining 
of the velocity-space structure. The physical agent of this coarse graining 
is collisions, even if they would appear to be infinitesimally small. Mathematically, 
the \cite{landau36} collision operator is a diffusion operator in velocity 
space and so even small collision frequencies are enough to thermalise 
any amount of energy, provided sufficiently large velocity-space gradients 
develop. 

In a {\em linear} plasma system, Landau damping, or, more generally, phase mixing, 
is the only available thermalisation route. It provides an adequate 
mechanism to process any injected free energy at any fixed wave number 
(since the process is linear, energy will stay in the wave number into 
which it is injected; there is no coupling), leading to a finite effective 
damping rate and filling up the phase space with free energy. If one uses a Hermite 
decomposition to quantify ``scales'' in velocity space, one finds that, in a steady-state system 
continuously pumped via low Hermite moments and dissipating free energy via high ones, 
the free energy will accumulate in phase space to a level that diverges 
if the collisionality is taken to zero; the collisional heating rate in this 
limit is finite and equal to the phase-mixing rate \citep{kanekar15}. 
How does this mechanism coexist and compete with the refinement 
of spatial scales caused by coupling between scales---a well-nigh inevitable 
consequence of nonlinearity?  

In this paper, we address this question using a simple archetypal example 
of plasma turbulence---electrostatic turbulence in a drift-kinetic 
plasma. We will describe this example in \secref{sec:prelims}, along with 
all the relevant preliminaries: the concept of free energy, the Hermite decomposition, 
and the existing Kolmogorov-style ``fluid'' turbulence theory for this 
problem \citep{barnes11}. 
In \secref{sec:formalism}, we will introduce the phase-space formalism that explicitly 
separates the phase-mixing and the ``anti-phase-mixing'' perturbations (the latter 
activated by the plasma echo effect), 
both of which turn out to be inevitable in a nonlinear system, and provides 
a useful starting point for a substantive theoretical treatment of phase-space 
turbulence. In \secref{sec:scaling}, a phenomenological scaling theory 
of this turbulence will be proposed. While we will describe in detail how 
free energy and its fluxes are distributed in the inertial range---leading to some 
interesting and testable scalings---the main conclusion will be that 
phase mixing is quite heavily suppressed in a turbulent system. 
\Secref{sec:concs} is devoted to summarising this and other findings 
and to discussing their implications, as well as future directions 
of travel. A reader only interested in a digest can skip to this section now. 
 
\section{Preliminaries}
\label{sec:prelims}

This section contains a rather extended tutorial on a number of topics 
constituting elementary but necessary background to what will follow. 
Readers who are sufficiently steeped in these matters 
can skim through this section and then dedicate themselves more seriously to 
\secsand{sec:formalism}{sec:scaling} (where references to relevant 
parts of \secref{sec:prelims} will be supplied). 

\subsection{Prototypical kinetic problem}

We consider a plasma near Maxwellian equilibrium, in which case  
the distribution function for particles of species $s$ can be expressed as 
\beq
f_s = F_{{\rm M}s} + \df_s,
\eeq
where $F_{{\rm M}s}$ is a Maxwellian distribution and $\df_s$ a small 
perturbation. 

We assume this plasma to be in a uniform strong magnetic field $\vB=B\vz$ 
($\vz$ is the unit vector in the direction of this field, designated the $z$ axis). 
We consider low-frequency perturbations, which will be highly anisotropic 
with respect to the field: 
\beq
\omega \ll \Omega_s,\quad
\kpar \ll \kperp, 
\eeq
where $\Omega_s$ is the Larmor frequency.

We assume these perturbations to be electrostatic, viz., 
\beq
\delta\vE = -\vdel\phi,\quad
\delta\vB = 0,
\eeq
where $\phi$ is the scalar potential. We use Gaussian electromagnetic units. 

We consider only long wavelengths, 
\beq
\kperp\rho_s \ll 1,
\eeq
where $\rho_s$ is the Larmor radius. 

Finally, we assume a Boltzmann electron response (which arises via expansion of the 
electron drift-kinetic equation in the electron-to-ion mass ratio):\footnote{The intricacies 
of the $\kpar=0$ electron response (see \secref{sec:nuance}) do not affect 
the inertial-range theory to be presented here.}
\beq
\label{eq:Boltz}
\frac{e\phi}{T_e} = \frac{\dn_e}{n_e} 
= \frac{\dn_i}{n_i} =\frac{1}{n_i}\int\rmd^3 v\, \df_i,
\eeq
where $e$ is the electron charge, $T_s$ and $n_s$ are the equilibrium 
temperatures and number densities, respectively, and $\dn_s$ are density 
perturbations ($s=e$ for electrons, $s=i$ for ions). 
The second equality in \eqref{eq:Boltz} is a consequence of plasma quasineutrality. 
It is useful to denote
\beq
\ephi = \frac{Ze\phi}{T_i},
\eeq
where $Z$ is the ratio of the ion to electron charge. 

Under these assumptions, we may integrate out the dependence of the ion 
distribution function on perpendicular velocities, so we introduce 
\beq
g(t,\vr,\vpar) = \frac{1}{n_i}\int\rmd^2\vvperp \df_i,
\eeq
and write the drift-kinetic equation for $g$ in a 4D phase space:  
\begin{align}
\label{eq:g}
\frac{\dd g}{\dd t} + &\vpar\dpar(g + \ephi\FM) + \vuperp\cdot\vdperp g = C[g] + \chi,\\
&\ephi = \alpha\int\rmd\vpar g,\quad \alpha = \frac{ZT_e}{T_i},
\label{eq:phi_alpha}
\end{align}
where $\FM$ is a 1D Maxwellian with thermal speed $\vth$,
\beq
\FM = \frac{1}{\sqrt{\pi}}\,e^{-\vpar^2/\vth^2},\quad
\vth = \sqrt{\frac{2T_i}{m_i}},
\eeq
$\vuperp$ is the $\vE\times\vB$ drift velocity,
\beq
\vuperp = c\,\frac{\delta\vE\times\vB}{B^2} = \frac{\rho_i\vth}{2}\,\vz\times\vdperp\ephi,
\label{eq:uperp}
\eeq
$C[g]$ is the collision operator and $\chi$ a source term---both of which need 
a little further discussion, which we will provide in \secref{sec:inj_etc}. 

Note that while we will be referring to ``slab'' ion-temperature-gradient (ITG) 
turbulence \citep[e.g.,][]{cowley91,ottaviani97,horton99} as the main physical instantiation 
that we have in mind of the kinetic problem described above, there will be nothing 
in our theory that would make it inapplicable to the (inertial range of) 
electron-temperature-gradient (ETG) turbulence \citep{dorland00,jenko00}, 
or indeed to a generic case of electrostatic drift-kinetic turbulence with 
energy injection at long wavelengths. 

\subsubsection{A nuance: Boltzmann closure and zonal flows}
\label{sec:nuance}

In this context, 
we must come clean on an important detail. The Boltzmann closure \exref{eq:Boltz}
for the electron density is, in fact, only valid for perturbations with $\kpar\neq0$
because it relies on electrons streaming quickly along the magnetic field lines to short out 
the parallel electric field. In tokamak plasmas, where magnetic shear imposes a link 
between $\kpar$ and $k_y$, the Boltzmann closure is normally amended 
\citep{dorland93,hammett93} 
to remove from the electron density the response associated with perturbations 
that have $k_y=\kpar=0$ (the ``zonal flows''), namely, 
\beq
\frac{\dn_e}{n_e} = \frac{e(\phi-\overline{\phi})}{T_e},
\eeq 
where $\overline{\phi}$ is the flux-surface average, which in our context is an average 
over $y$ and $z$ (formally, one gets this by first deriving the density response 
in a toroidal, magnetically sheared system, as is done, e.g., in \S J.2 of \citealt{Abel13b}, 
then taking the magnetic shear and curvature to be small and passing to the slab limit). 

This implies that, in order to find the $(y,z)$-averaged
(zonal) part of $\ephi$ from the ion distribution 
function, at least the lowest-order finite-Larmor-radius correction has to be kept
(physically representing the polarisation drift; see \citealt{krommes93,krommes10}), 
leading to $\alpha$ in \eqref{eq:phi_alpha} for the zonal part of $\ephi$ having 
to be replaced with $\alpha=2/k_x^2\rho_i^2$, or
\beq 
-\frac{1}{2}\,\rho_i^2\dd_x^2\overline{\ephi} = \int\rmd\vpar\overline{g}.
\eeq 
In ITG turbulence far from marginal stability, these changes 
affect important quantitative details of the interaction between zonal flows 
and drift waves at the outer scale (\citealt{rogers00}; see also discussion around \eqref{eq:S_ZF}), 
but do not matter for the inertial-range physics that we will focus on in this paper. 

A reader who is unconvinced may observe 
that \eqref{eq:phi_alpha} can be used without these modifications if, instead of 
considering ITG turbulence, we consider ETG turbulence \citep{dorland00,jenko00}. 
In this case, it is the ions that have a Boltzmann response (due to their large Larmor orbits, 
over which the density response from electron-scale fluctuations averages out), 
\beq
\frac{\dn_i}{n_i} = - \frac{Ze\phi}{T_i} 
\eeq
(which is $=\dn_e/n_e$ by quasineutrality). 
The required modifications in \eqsand{eq:g}{eq:phi_alpha}
are 
\beq
\ephi\to \frac{e\phi}{T_e},\quad 
\alpha\to - \frac{T_i}{ZT_e},\quad 
\rho_i\to\rho_e,\quad
\vth\to\vthe=\sqrt{\frac{2T_e}{m_e}},
\eeq
and $\ephi\FM\to-\ephi\FM$ in \eqref{eq:g}. None 
of this affects anything essential in the upcoming theoretical developments.

\subsubsection{Injection, phase mixing, advection, dissipation} 
\label{sec:inj_etc}

The precise nature of the source term $\chi$ in \eqref{eq:g} will not matter in our theory, 
as long as it does not contain any sharp dependence on $\vpar$ (i.e., is confined 
to low velocity moments). 
A random forcing is often a convenient choice for analytical theory 
\citep[e.g.,][]{plunk13,plunk14,kanekar15}, but a more physical form 
in the context of electrostatic drift-kinetic turbulence in plasmas \citep[e.g.,][]{horton99}
arises from accounting for the presence of equilibrium density and temperature gradients, 
taken, conventionally, to be in the negative $x$ direction:\footnote{The erudite reader 
given pause by $1/2$ rather than $3/2$ in the prefactor of $1/L_T$ in \eqref{eq:chi}
will recall that we have integrated out the $\vperp$ dependence.} 
\begin{align}
\label{eq:chi}
\chi = -\frac{\vuperp\cdot\vdel (n_i \FM)}{n_i} 
= - \frac{\rho_i\vth}{2}\frac{\dd\ephi}{\dd y}
\lt[\frac{1}{L_n} + \lt(\frac{\vpar^2}{\vth^2}-\frac{1}{2}\rt)\frac{1}{L_T}\rt]\FM,\\
\frac{1}{L_n} = -\frac{1}{n_i}\frac{\rmd n_i}{\rmd x},
\quad
\frac{1}{L_T} = -\frac{1}{T_i}\frac{\rmd T_i}{\rmd x}.
\nonumber
\end{align}
We shall see in \secref{sec:Hermite} that these terms render the system linearly 
unstable and thus extract energy from the equilibrium gradients and inject it 
into the perturbed distribution. 

The resulting perturbations are subject to two influences, linear and nonlinear, 
encoded by the second ($\vpar\dpar g$) and fourth ($\vuperp\cdot\vdperp g$) 
terms on the left-hand side of \eqref{eq:g}, respectively. 
The nonlinear term represents advection of the distribution function by 
the mean perpendicular flow, itself determined by the former. This 
involves coupling between different wave numbers and thus usually leads to 
spatial mixing (generation of small spatial scales) of the perturbed distribution.  
The linear term represents phase mixing---generation of small velocity-space scales 
in the perturbed distribution function. The simplest way to understand this is 
to notice that the homogeneous solution to the linear kinetic equation in Fourier 
space, $\dd_t g + i\vpar\kpar g = \dots$, is $g\sim e^{-i\vpar\kpar t}$ and the 
velocity gradient of that grows secularly with time, $\dd_{\vpar} g = -i\kpar t g$.  

As fine structure in phase space is generated, there must be a means for removing it. 
This is why, even for a ``collisionless'' (meaning in fact weakly collisional) plasma, 
the collision operator $C[g]$ must be included in \eqref{eq:g}. 
We hasten to acknowledge that, in pretending that the collision operator operates 
purely on $g$, we are ignoring that the $\vvperp$ dependence cannot 
in fact be integrated out of it: collisions will strive to isotropise the distribution and 
so the collision operator must necessarily couple $\vvperp$ and $\vpar$. However, 
non-rigorously, when the collision frequency is small, 
\beq
\nu \ll \omega,\ \kpar\vth,\ \kperp\uperp,
\eeq
the collision operator's essential contribution will be simply to iron out fine structure 
in velocity space and, given an initial distribution and a source that 
are smooth in $\vv$, only fine structure in $\vpar$ can arise. 
Thus, it should suffice to assume a simple model form for $C[g]$: 
for example, the \cite{lenard58} operator, 
\beq
C[g] = \nu\frac{\dd}{\dd\hvpar}\lt(\frac{1}{2}\frac{\dd}{\dd\hvpar} + \hvpar\rt)g,
\quad \hvpar = \frac{\vpar}{\vth}.
\label{eq:LB}
\eeq
The fact that this operator does not conserve momentum or energy, while easily 
repaired if one strives for quantitatively precise energetics \citep{kirkwood46}, 
will not cause embarrassment as collisions will only matter for high velocity 
moments (because large gradients with respect to $\vpar$ are necessary to 
offset the smallness of $\nu$). It is not hard to estimate the velocity-space
scales at which collisions can become important: balancing $C[g]\sim\omega g$,
where $\omega\sim\kpar\vth$ and/or $\kperp\uperp$ 
is the typical frequency scale of the collisionless dynamics, we find that 
the requisite velocity scale is 
\beq
\frac{\delta\vpar}{\vth}\sim \lt(\frac{\nu}{\omega}\rt)^{1/2},
\eeq 
so the structure gets ever finer as $\nu\to+0$. 

Finally, we are going to assume implicitly that \eqref{eq:g} contains some regularising 
term to ensure a cutoff in $\kperp$---a necessity because of the spatial 
mixing associated with the nonlinear advection. Physically, the advection term 
will drive the system out of the domain of validity of the drift-kinetic approximation, 
to $\kperp\rho_i\sim1$ and larger. The precise way in which the energy is thermalised 
at these Larmor and sub-Larmor scales is a rich and interesting topic in its own right, 
involving a kinetic cascade in a 5D phase space (with nonlinear phase mixing 
in $\vperp$ now also occurring)---but these matters are outside the scope of 
this treatment \citep[see][]{sch08,tome,tatsuno09,plunk10,banon11prl}. 

It is the competition between the two ways---linear phase mixing vs.\ nonlinear advection---of 
generating small-scale structure in phase space and thus enabling 
the energy injected by the source to be thermalised that will be the subject of this paper. 

\subsection{Free energy}

We have referred to injection and thermalisation of energy many times now, and so 
defining precisely what we mean by ``energy'' has become overdue. 

Energy in $\df$ kinetics 
(i.e., in near-equilibrium kinetics) is the free energy associated with the perturbed 
distribution:\footnote{The understanding that this is the case can be traced back 
through a sequence of papers, from early, somewhat forgotten, insights to a more 
recent surge in appreciation \citep{kruskal58,bernstein58,fowler63,fowler68,krommes94,krommes99,sugama96,hallatschek04,howes06,candy06,sch08,tome,scott10,banon11prl,banon11,abel13,kunz15,parker15}. Note that we have not included in \eqref{eq:FF} the energy of the electric and magnetic field,  
$(\la E^2\ra + \la \delta B^2\ra)/8\pi$ (which is part of the general expression for the 
free energy; see, e.g., 
\citealt{sch08}) because we are considering electrostatic perturbations ($\delta B=0$) 
at scales much longer than the Debye length ($\la E^2\ra$ is negligible).} 
\beq
{\cal F} = -\sum_s T_s\delta S_s = 
\sum_s T_s\,\delta\!\!\int\rmd^3\vv\la f_s\ln f_s\ra 
= \sum_s \int\rmd^3\vv\frac{T_s\la\df_s^2\ra}{2 F_{\mathrm{M}s}},
\label{eq:FF}
\eeq
where angle brackets denote spatial averaging and 
$\delta S_s$ is the mean additional (negative!) entropy associated with 
the perturbed distribution of species $s$. The last expression in \eqref{eq:FF} 
was obtained by letting $f_s = F_{\mathrm{M}s} + \df_s$ and expanding $\la f_s\ln f_s\ra$
to second order in $\df_s$ (see, e.g., \citealt{sch08}; note that $\la\df_s\ra=0$ 
because $\la f_s\ra = F_{\mathrm{M}s}$ by definition). 
It is now not hard to establish that ${\cal F} = n_i T_i W$, where 
\beq
W = \int\rmd\vpar \frac{\la g^2\ra}{2\FM} + \frac{\la\ephi^2\ra}{2\alpha} 
\label{eq:W}
\eeq
is the quadratic quantity conserved by \eqref{eq:g}. This can be 
shown either by using the Boltzmann-electron closure in \eqref{eq:FF} 
(viz., $\df_e=(e\phi/T_e)F_{\mathrm{M}e}$, so the $s=e$ term in $\sum_s$
gives rise to the $\la\ephi^2\ra$ term in $W$) 
or directly starting from \eqref{eq:g}, which gives us the following 
law of evolution of the free energy: 
\beq
\frac{\rmd W}{\rmd t} = \int\rmd\vpar\lt(\frac{\la g\chi\ra}{\FM} + \la\ephi\chi\ra\rt) 
+ \int\rmd\vpar \frac{\la g C[g]\ra}{\FM}. 
\eeq 
The first term on the right-hand side is the energy-injection term, 
which turns into the usual flux term for ITG (or ETG) turbulence if we substitute 
$\chi$ from \eqref{eq:chi} (see \eqref{eq:Wfluid} below), 
and the second, negative definite, term 
is the collisional thermalisation of this energy flux. 

The Landau damping of the electrostatic perturbations is simply the 
transfer of free energy, via phase mixing, from the $\la\ephi^2\ra$ part 
of $W$ to $\la g^2\ra$\footnote{To be precise, from $\la\ephi^2\ra$ and 
low-order (``fluid'') velocity moments of $g$ to higher-order (``kinetic'') 
moments (see \secref{sec:fe_flows}).}: 
since $\ephi = \alpha\int\rmd\vpar g$, 
small-scale velocity-space structure in $g$ is washed out in $\ephi$ 
but of course remains as free energy in~$\la g^2\ra$ 
\citep[cf.][]{hammett90,hammett92}. 

\subsection{Hermite decomposition}
\label{sec:Hermite}

A natural way to separate the ``fluid'' part of the problem from 
the ``kinetic'' one and represent phase mixing is to expand the perturbed 
distribution in Hermite polynomials:\footnote{This has attracted recurring 
bursts of attention over many years, especially recently \citep{grad49,armstrong67,grant67a,eltgroth74,crownfield77,hammett93,parker95,smith97,ng99,watanabe04,zocco11,black13,loureiro13,hatch13,hatch14,plunk14,kanekar15,kanekar15phd,parker15}.}
\begin{align}
g(\vpar) &= \sum_{m=0}^\infty \frac{H_m(\hvpar)\FM(\vpar)}{\sqrt{2^m m!}}\,g_m,\\
g_m &= \int\rmd\vpar \frac{H_m(\hvpar)}{\sqrt{2^m m!}}\,g(\vpar),
\end{align}   
where $\hvpar=\vpar/\vth$ and the ``physicist's'' Hermite polynomials are
\beq
H_m(\hvpar) = (-1)^m e^{\hvpar^2}\frac{\rmd^m}{\rmd\hvpar^m}\, e^{-\hvpar^2},
\quad
\int\rmd\vpar \frac{H_m(\hvpar) H_n(\hvpar)}{2^m m!}\,\FM(\vpar) = \delta_{mn}.
\eeq
The first three Hermite moments are the (ion) density ($\dn$), mean-parallel-velocity ($\upar$)
and parallel-temperature ($\dTpar$) perturbations: 
\begin{align}
H_0(\hvpar) = 1
\quad&\Rightarrow\quad
g_0 = \frac{\dn}{n} = \frac{\ephi}{\alpha},\\
H_1(\hvpar) = 2\hvpar
\quad&\Rightarrow\quad
g_1 = \sqrt{2}\,\frac{\upar}{\vth},\\
H_2(\hvpar) = 4\lt(\hvpar^2 - \frac{1}{2}\rt)
\quad&\Rightarrow\quad
g_2 = \frac{1}{\sqrt{2}}\frac{\dTpar}{T}.
\end{align}
Noting further that the source term, \eqref{eq:chi}, is 
\beq
\chi = -\frac{\rho_i\vth}{2}\frac{\dd\ephi}{\dd y}\lt[\frac{H_0(\hvpar)}{L_n} 
+ \frac{H_2(\hvpar)}{4L_T}\rt]\FM
\equiv \lt[\chi_0 + \frac{H_2(\hvpar)}{2\sqrt{2}}\,\chi_2\rt]\FM
\eeq
and that the streaming term in \eqref{eq:g}, $\vpar\dpar g$, couples 
Hermite moments of adjacent orders via the formula 
\beq
\hvpar H_m(\hvpar) = \frac{1}{2}\,H_{m+1}(\hvpar) + mH_{m-1}(\hvpar),
\eeq
we arrive at the following Hermite representation of \eqref{eq:g}:
\begin{align}
\label{eq:phi}
&\quad\!\frac{\dd}{\dd t}\frac{\ephi}{\alpha} + \vth\dpar\frac{\upar}{\vth} 
= \chi_0 = -\frac{\vth}{2L_n}\rho_i\frac{\dd\ephi}{\dd y},\\
\label{eq:upar}
&\lt(\frac{\dd}{\dd t} + \vuperp\cdot\vdperp\rt)\frac{\upar}{\vth} 
+ \vth\dpar\lt(\frac{1}{2}\frac{\dTpar}{T} + \frac{1+\alpha}{\alpha}\,\ephi\rt) = 0,\\
\label{eq:dTpar}
&\lt(\frac{\dd}{\dd t} + \vuperp\cdot\vdperp\rt)\frac{\dTpar}{T} 
+ \vth\dpar\lt(\sqrt{3}\,g_3 + 2\frac{\upar}{\vth}\rt) = 
\sqrt{2}\,\chi_2 = -\frac{\vth}{2L_T}\rho_i\frac{\dd\ephi}{\dd y},
\end{align}
and, for $m\ge3$, a universal equation retaining no traces of the temperature-gradient 
drive or Boltzmann-electron physics:
\beq
\lt(\frac{\dd}{\dd t} + \vuperp\cdot\vdperp\rt) g_m + 
\vth\dpar\lt(\sqrt{\frac{m+1}{2}}\,g_{m+1} + \sqrt{\frac{m}{2}}\,g_{m-1}\rt)
= -\nu m g_m. 
\label{eq:gm}
\eeq
Note that we have taken advantage of the fact that Hermite polynomials are eigenfunctions 
of the Lenard--Bernstein operator \exref{eq:LB}, but ignored collisions in the 
$m=1$ and $m=2$ equations (this is allowed because we are assuming $\nu\to+0$ and so 
collisions will only be important at $m\gg1$). 

\subsubsection{Energy injection: slab ITG instability}

The first three equations are the standard three-field fluid system that 
describes an ITG-unstable plasma at long wavelengths in an unsheared slab \citep{cowley91}. 
The quickest way to obtain the slab ITG instability \citep{rudakov61,coppi67,cowley91} 
is to balance the two terms 
on the left-hand side of \eqref{eq:phi}, the first with the third term 
in \eqref{eq:upar}, and the first term on the left-hand side with the temperature-gradient 
term on the right-hand side of \eqref{eq:dTpar}. The resulting dispersion 
relation has three roots, of which one is unstable: 
\beq
\omega^3 \approx \frac{\alpha}{2}(\kpar\vth)^2\omega_{*T}
\quad\Rightarrow\quad
\omega \approx \lt(-\frac{1}{2}+i\,\frac{\sqrt{3}}{2}\rt)
\lt(\frac{\alpha}{2}\rt)^{1/3}(\kpar\vth)^{2/3}\omega_{*T}^{1/3},
\eeq
where $\omega_{*T}=k_y\rho_i\vth/2L_T$. This approximation is valid provided 
$L_n/L_T\gg1$ and $\omega_{*T}\gg \kpar\vth$, although, as the growth rate 
grows with $\kpar$, the fastest growth is in fact achieved for 
$\kpar\vth\sim\omega_{*T}$, when the dispersion relation is a more complicated 
and somewhat unedifying equation. At $\kpar\vth\gg\omega_{*T}$, the 
ITG mode is replaced by a sound wave, which, in a kinetic system, 
is heavily Landau damped.\footnote{An elementary analysis of the slab ITG dispersion relation
can be found, e.g., in Appendix B.2 of \citet{sch12}. Note that, in \secref{sec:CB}, we will 
argue that the inertial-range fluctuations in fact have $\kpar\vth\gg\omega_{*T}$, 
and in \secref{sec:scaling}, we will show that their Landau damping is suppressed 
in the nonlinear regime.} 

\subsubsection{Free-energy flows}
\label{sec:fe_flows}

The temperature-gradient instability injects energy into the $\ephi$, $\upar/\vth$ and $\dTpar/T$ 
perturbations, all of which are comparable to each other in magnitude when  
$\kpar\vth\sim\omega_{*T}$. Because the three-field system is not closed,\footnote{The only 
rigorous way to turn it into a closed system is to assume $\nu\gg\omega$, $\kpar\vth$, 
$\kperp\uperp$ in \eqref{eq:gm}, whence $g_m\gg g_{m+1}$ and so the heat flux is expressible 
in terms of the temperature gradient:  
$\sqrt{3}\,g_3\approx (\vth/\sqrt{2}\,\nu)\dpar g_2 = (\vth/2\nu)\dpar\dTpar/T$. 
Putting this into \eqref{eq:dTpar} gives rise to a parallel heat conduction term. 
We are not, however, interested in this collisional limit.} 
there is a transfer of energy from $\dTpar/T$ to higher Hermite moments: 
the $g_3$ term in \eqref{eq:dTpar} provides the energy sink from the 
unstable (``forced'') moments and the $g_2$ term in \eqref{eq:gm} at $m=3$ 
is the source for the higher moments; the energy thus received by them is 
eventually thermalised via collisions. 

To be more precise about these statements, let us rewrite the free energy 
\exref{eq:W} in terms of Hermite moments: 
\beq
W = \frac{1+\alpha}{2\alpha^2}\la\ephi^2\ra + \frac{\la\upar^2\ra}{\vth^2} + 
\frac{1}{4}\frac{\la\dTpar^2\ra}{T^2} + \frac{1}{2}\sum_{m=3}^\infty\la g_m^2\ra.
\eeq
Its ``fluid'' and ``kinetic'' parts satisfy:\footnote{Note that 
both \eqsand{eq:Wfluid}{eq:Wkin} will also contain sinks accounting 
for energy losses at small spatial scales.} 
\begin{align}
\label{eq:Wfluid}
&\frac{\rmd}{\rmd t}\lt(\frac{1+\alpha}{\alpha^2}\la\ephi^2\ra + \frac{\la\upar^2\ra}{\vth^2} + 
\frac{1}{4}\frac{\la\dTpar^2\ra}{T^2}\rt) 
= \frac{\la\dTpar u_x\ra}{2TL_T} - \frac{\sqrt{3}\,\vth}{2 T}\la\dTpar\dpar g_3\ra,\\
\label{eq:Wkin}
&\frac{\rmd}{\rmd t}\frac{1}{2}\sum_{m=3}^\infty\la g_m^2\ra
= \frac{\sqrt{3}\,\vth}{2 T}\la\dTpar\dpar g_3\ra - \nu\sum_{m=3}^\infty m\la g_m^2\ra. 
\end{align}
The first term on the right-hand side of \eqref{eq:Wfluid} is the injected energy flux  
and the second term on the right-hand side of \eqref{eq:Wkin} is the dissipation of that 
flux by collisions. In steady state, $\rmd \la\dots\ra/\rmd t=0$, 
we must have 
\beq
\la\dTpar\dpar g_3\ra \ge 0
\label{eq:transfer}
\eeq
because the collision term is negative-definite 
in \eqref{eq:Wkin}, and, therefore, 
\beq
\la\dTpar u_x\ra \ge 0
\eeq
to achieve balance in \eqref{eq:Wfluid}. 
The inequality \exref{eq:transfer} implies a non-negative mean energy flux 
to higher Hermite moments \citep[cf.][]{krommes94,nakata12}. 

How that flux is processed from being injected at $m=3$ to being dissipated at $m\gg1$
(assuming $\nu\to+0$) is handled by \eqref{eq:gm}. 
This equation contains in a beautifully explicit form
the two effects to which this paper is devoted: 
the phase mixing is manifest in that $g_m$ is coupled to $g_{m+1}$ and $g_{m-1}$, 
providing a mechanism for pushing energy to higher $m$'s; simultaneously,  
all Hermite moments $g_m$ are advected (spatially mixed towards smaller scales) 
by the same fluctuating velocity $\vuperp$, determined, via \eqref{eq:uperp}, 
by the zeroth Hermite moment, $\ephi=\alpha g_0$.   

\subsection{``Fluid'' turbulence theory}
\label{sec:fluid_theory}

\citet{barnes11} proposed a Kolmogorov-style theory of ITG turbulence,
essentially ignoring the possibility of a leakage of free energy from the low 
Hermite moments to the high. While their theory is by no means 
uncontroversial or the only offering on the market \citep[e.g.,][]{gurcan09,plunk15}, 
it does appear to match the results of numerical experiments (in the strongly 
unstable regime) and so it is worth both reviewing how it is constructed 
and examining to what extent it contradicts the statement made 
in the previous subsection that free energy must leak to higher Hermite moments. 

The scaling argument of \citet{barnes11} addresses two main questions (as would any 
such argument aspiring to be a complete theory): 

(i) what is the effective 
outer (energy-containing) scale of the turbulence and the fluctuation level 
at that scale;

(ii) what is the spatial structure of the turbulence in the 
``inertial range'' between that outer scale and the small-scale cutoff?  

\subsubsection{Outer scale}
\label{sec:outer}

The first question would be trivial for turbulence forced 
externally at some fixed scale, but for temperature-gradient-driven turbulence, 
relying on a linear instability, the system must decide where to have its energy-containing 
scale. This is not simply the peak scale of the growth rate, because 
at long wave lengths the growth rate will generally grow with wave number 
less quickly than will the nonlinear cascade rate (as we shall see in \secref{sec:in_range})
and so in fact it is the largest scale at which there is an instability that will 
end up being the energy-containing scale. 

\citet{barnes11} conjectured that this infrared cutoff will be set by the 
largest {\em parallel} scale available to fluctuations: 
\beq
\kparo \sim \frac{1}{\Lpar},
\eeq
where $\Lpar$ in our idealised homogeneous system is simply the parallel 
extent of the ``box''---in a tokamak, it would be the magnetic connection 
length between unstable and stable parts of the plasma 
($\Lpar\sim qR$, where $q$ is safety factor and $R$ the major radius). 
The perpendicular energy-containing scale is then given~by 
\beq
\omega_{*T}=\kyo\rho_i\frac{\vth}{L_T}\sim\kparo\vth
\quad\Rightarrow\quad 
\kyo\rho_i \sim \frac{L_T}{\Lpar}
\label{eq:kyo}
\eeq
because the instability would be supplanted by stable (in fact, Landau-damped)
sound waves at smaller $k_y$. Note that we require $L_T/\Lpar\ll1$ 
for the turbulence to occur in a scale range consistent with the drift-kinetic 
approximation $\kperp\rho_i\ll1$. Finally, it is further conjectured that 
the zonal flows generated by the turbulence will have a typical 
shearing rate $S_{\mathrm{ZF}}$ comparable to the nonlinear 
decorrelation rate $\tauco^{-1}$ at the outer scale \citep[cf.][]{rogers00}
and, therefore, will isotropise the turbulence:\footnote{It is possible to 
imagine (or conjecture) variants of drift-wave turbulence in which zonal flows are not 
strong enough to do this. In such systems, the saturated state at the outer scale 
is dominated by ``streamers,'' anisotropic structures with $\kxo\ll\kyo$, whose 
radial extent is probably determined by the size of the system 
\citep{drake88,drake91,cowley91,rogers98,dorland00,jenko00}. In order for these 
structures to survive, they must be immune to the secondary instability 
that would otherwise give rise to zonal flows, which 
would in turn break up the streamers \citep{rogers00,quinn13,connaughton14}. 
How a streamer-dominated outer-scale state channels its energy 
into an inertial-range cascade is not entirely well understood. 
However, we do not expect that the physics of this inertial range 
to be much different from that described below.} 
\beq
\kxo \sim S_{\mathrm{ZF}}\kyo\tauco \sim \kyo \sim \kperpo. 
\label{eq:S_ZF}
\eeq 
This is the only place in the theory where the zonal flows make an appearance 
as it is assumed that they do not completely dominate the nonlinear dynamics, 
in contrast to their alleged behaviour in the near-threshold regime 
\citep[e.g.,][]{dimits00,diamond05,diamond11,gurcan09,nakata12,ghim13,connaughton14,makwana14}. 

The energy-containing scale given by \eqref{eq:kyo}, the amount of energy it contains 
is estimated by balancing the rate of injection by instability, $\omega_{*T}$, 
against the rate $\tauco^{-1}$ of nonlinear removal of this energy to smaller scales 
via advection by the turbulent flow:  
\beq
\omega_{*T}\sim \kperpo\rho_i\frac{\vth}{L_T}\sim\tauco^{-1}\sim\kperpo\uperpo 
\sim \rho_i\vth\kperpo^2\ephio
\quad\Rightarrow\quad
\ephio\sim\frac{1}{\kperpo L_T}\sim\frac{\rho_i\Lpar}{L_T^2}. 
\label{eq:phio}
\eeq

Finally, if one's overriding practical concern is the calculation of the 
effective heat transport caused by the turbulence, one concludes from the above 
that the turbulent thermal diffusivity and the heat flux are 
\begin{align}
\DT \sim \uperpo^2\tauco \sim \frac{\uperpo}{\kperpo} \sim \rho_i\vth\ephio,
\quad\Rightarrow\quad
Q \sim \frac{n\DT T}{L_T} \sim n\rho_i^2\vth\,\frac{\Lpar}{L_T^3}.  
\label{eq:Q}
\end{align}

All of this is not particularly sensitive 
to the fact that, in making the argument that led to \eqref{eq:phio}, we 
completely ignored the possibility (in fact, the inevitability) that some of 
the energy injected at the outer scale might be removed not by the nonlinear advection, 
as if the system were purely fluid, but also by the phase mixing towards high $m$'s. 
The presence of such a transfer (of which there is, in fact, numerical 
evidence; see, e.g., \citealt{watanabe06,hatch11prl,hatch11pop,nakata12}) would only break 
our argument if the rate $\sim\kparo\vth$ of this transfer were substantially larger 
than the nonlinear advection rate and so if the dominant balance were 
$\omega_{*T}\sim\kparo\vth\gg\kperpo\uperpo$. But this is obviously impossible 
as one cannot saturate a linear instability by a linear mechanism: there 
would not be anything in the theory to determine the saturated 
amplitude.\footnote{Again, focusing on turbulence far above the threshold, 
we are going to ignore the possibility of a more sophisticated scheme
involving zonal flows.} In view of \eqref{eq:kyo}, the phase mixing rate 
is, in fact, of the same order as both $\omega_{*T}$ and $\kperpo\uperpo$. 
Therefore, it cannot affect the basic scalings---although for the purposes 
of quantitative transport modelling, it is quite crucial to know by what 
fraction of order unity it might cut the nonlinear mixing rate, a 
key preoccupation in the development of ``Landau-fluid'' closures for plasma 
turbulence in fusion contexts 
\citep{hammett92,hammett93,dorland93,beer96,snyder01gf}. 

A question that is much more sensitive to whether phase mixing is 
nonnegligible is the structure of the inertial range. 

\subsubsection{Inertial range: perpendicular spectrum}
\label{sec:in_range}

How is the energy injected at $(\kperpo,\kparo)$ cascaded to smaller scales? 
Ignoring phase mixing, \citet{barnes11} proposed to calculate the dependence 
of the turbulent amplitudes on scale via the Kolmogorov assumption of 
constant energy flux: at each scale $\kperp^{-1}$, energy $\ephi^2$ is 
transferred (locally) to the next smaller scale over the cascade time~$\tauc$: 
\beq
\frac{\ephi^2}{\tauc} \sim \kperp\uperp\ephi^2 \propto \kperp^2\ephi^3 = \const
\quad\Rightarrow\quad
\ephi \propto \kperp^{-2/3},  
\label{eq:constflux}
\eeq
where we used $\uperp\propto\kperp\ephi$ (see \eqref{eq:uperp}). 
Note that, both here and in similar arguments that will follow, we do not make a 
distinction between the energy content of low-$m$ moments, 
assuming\footnote{This is because the typical rate for coupling these 
moments is $\kpar\vth$, which will shortly be argued to be comparable 
to the nonlinear rate at which these moments change, \eqref{eq:CB}.} 
\beq
\ephi \sim \frac{\upar}{\vth} \sim \frac{\dTpar}{T}
\eeq
and possibly also $\sim$ a few more low-$m$ moments of $g$, although 
we do assume that there is not a substantial energy leakage to asymptotically 
large $m$'s. The 1D (perpendicular) spectrum is then 
\beq
\Eph^\perp(\kperp) = 2\pi\kperp\int\rmd\kpar\la|\ephi_{\vk}|^2\ra 
\sim \frac{\ephi^2}{\kperp}\propto \kperp^{-7/3}, 
\label{eq:Ephi1D}
\eeq
where $\la\cdots\ra$ now denotes a time or ensemble average. 
This scaling is supported both by numerical simulations of \citet{barnes11} and, 
apparently, by those done by other groups 
\citep[][who confirm finding the same scaling, without, however, providing plots]{hatch13,hatch14,plunk15}. 

\subsubsection{Critical balance} 
\label{sec:CB}

The structure of the turbulence in the parallel direction can now 
be inferred via a causality argument known in the astrophysical MHD literature 
as ``critical balance'' \citep{GS95,GS97,boldyrev05} and 
emerging as a universal scaling principle for strong turbulence in 
wave-supporting systems \citep{cho04,tome,nazarenko11}: fluctuations 
cannot stay correlated at parallel distances longer than those over 
which linear communication happens at the same rate as the nonlinear decorrelation: 
thus, fluctuations are uncorrelated for 
\beq
\kpar\vth\lesssim\kperp\uperp \propto \kperp^{4/3}
\quad\Rightarrow\quad
\kpar\Lpar \lesssim\lt(\frac{\kperp}{\kperpo}\rt)^{4/3}. 
\label{eq:kpar_ineq} 
\eeq
Here and in what follows, we shall adopt a nondimensionalisation 
$\Lpar=1$ and $\kperpo=1$, so the above condition will henceforth 
be written $\kpar\lesssim\kperp^{4/3}$. 

This argument implies that, at any given $\kperp$, the ``energy-containing'' 
parallel scale will be given by the ``critical-balance'' wave number:  
\beq
\kparc\vth\sim\kperp\uperp
\quad\Rightarrow\quad
\kparc\sim\kperp^{4/3}, 
\label{eq:CB}
\eeq
another scaling that was confirmed numerically by \citet{barnes11}. 
The consequent scaling of the 1D parallel spectrum is 
(using \eqref{eq:CB} in \eqref{eq:constflux})\footnote{Another way 
of arriving at this spectrum and at the critical balance \citep{beresnyak15} 
is to start with the constant-flux conjecture applied to the scaling of 
amplitudes with {\em frequencies}, rather than wavenumbers: 
$\ephi^2\omega\sim\const\Rightarrow\ephi\propto\omega^{-1/2}$ \citep{corrsin63}. 
The frequencies of the perturbations will be $\omega\sim\kpar\vth$, hence
the parallel scaling \exref{eq:Epar1D}.}
\beq
\ephi \propto \kpar^{-1/2}
\quad\Rightarrow\quad
\Eph^\parallel(\kpar) = 2\pi\int\rmd\kperp\kperp\la|\ephi_{\vk}|^2\ra 
\sim \frac{\ephi^2}{\kpar}\propto \kpar^{-2}. 
\label{eq:Epar1D}
\eeq

Note that, under this scheme, the drift waves are slow in the inertial range 
because $\omega_{*T} \propto k_y$ whereas $\kperp\uperp\propto \kperp^{4/3}$,  
so the relevant frequency in \eqref{eq:kpar_ineq} is indeed $\sim\kpar\vth$, 
not $\omega_{*T}$. By the same token, energy injection by the temperature-gradient 
instability is slow compared to the nonlinear cascade rate, so, effectively, the instability 
only operates at the outer scale, while the fluctuations that carry the injected 
energy through the inertial range are more akin to ion sound waves than to drift waves.\footnote{This 
also explains why the \citet{barnes11} cascade should asymptotically override the nonlinear transfer 
proposed by \citet{gurcan09}: the latter authors argue, effectively, that the cascading 
of the energy to small scales is done by the nonlocal shearing of the drift 
waves by zonal flows, which they assume 
to occur at the rate $\sim\kperp u_\mathrm{ZF}$, where $u_\mathrm{ZF}$ is 
a scale-independent zonal velocity; this, via a constant-flux argument 
analogous to \exref{eq:constflux}, 
gives $\Eph^\perp\propto\kperp^{-2}$. However, if the zonal shearing 
rate is comparable to the 
energy-injection rate at the outer scale (which we also assume; 
see \eqsand{eq:S_ZF}{eq:phio}: $\kperpo u_\mathrm{ZF} \sim S_\mathrm{ZF} \sim 
\tauco^{-1}\sim\omega_{*T}$), then it will be smaller than $\kperp\uperp$ 
for $\kperp>\kperpo$. Note also that a nonlinear transfer rate $\propto\kperp$ 
could not effectively dominate the injection rate, $\omega_{*T}$, 
which is also $\propto\kperp$.} 

\subsubsection{Constant flux is inconsistent with robust phase mixing}
\label{sec:inconsistent}

In \secref{sec:scaling_phi}, we will explain how to derive from these arguments 
the scaling of the 2D spectra for any $\kperp$ and $\kpar$. However, we must first 
discuss the key point that the constant-flux assumption \exref{eq:constflux} cannot 
be consistent with both the idea that the energy resides along the 
``critical-balance curve'' \exref{eq:CB} {\em and} with 
phase mixing taking energy out to large $m$'s at the rate $\sim\kpar\vth$---simply 
because the latter would mean that the energy in the low-$m$ moments is not 
conserved and so need not be fully transferred nonlinearly to smaller scales. 

The simplest way to explain the implications of this for the spectra 
is to replace \eqref{eq:constflux} by a simple mock-up of an evolution equation 
for $\Eph^\perp(\kperp)$ \citep[cf.][]{batchelor53,howes08jgr}:  
\beq
\frac{\dd\Eph^\perp}{\dd t} = - \frac{\dd \eps}{\dd\kperp} -\gamma\Eph^\perp,
\quad
\eps \sim \frac{\kperp\Eph^\perp}{\tauc},
\quad
\gamma \sim \kpar\vth\sim\tauc^{-1}\sim\kperp^2\sqrt{\kperp\Eph^\perp}, 
\label{eq:batchelor}
\eeq
where $\eps$ is the energy flux and $\gamma$ is the effective rate of phase 
mixing (Landau damping), which, by the critical-balance conjecture \exref{eq:CB}, 
is of the same order as the cascade rate $\tauc^{-1}$. Assuming steady state 
in \eqref{eq:batchelor} and letting $\gamma\tauc=\xi=\const\sim1$ (independent 
of $\kperp$, as per critical balance), we get
\beq
\frac{\dd \eps}{\dd\kperp} = - \frac{\xi}{\kperp}\,\eps
\quad\Rightarrow\quad
\eps \propto \kperp^{-\xi}
\quad\Rightarrow\quad
\Eph^\perp(\kperp) \propto \kperp^{-(7+2\xi)/3}.  
\label{eq:Ephi_xi}
\eeq
Thus, the flux decreases with increasing wave number and so the spectrum is steeper 
than the constant-flux solution \exref{eq:Ephi1D}. The power laws that 
emerge in such dissipative systems are generally hard to predict and 
probably nonuniversal \citep[cf.][]{bratanov13prl,passot15}---in our case, 
because they depend on an order-unity prefactor ($\xi$) in 
the critical-balance relation \exref{eq:CB}, rather than on some 
dimensionally and physically inevitable scaling.\footnote{It is easy 
to see that $\xi<1$. Indeed, the nonlinear cascade rate 
that follows from \eqref{eq:Ephi_xi} is $\kperp\uperp\propto\kperp^{(4-\xi)/3}$, 
which can only overcome the injection rate associated with the temperature gradient 
if $\xi<1$ (see discussion at the end of \secref{sec:CB}). The extreme case 
$\xi=1$ gives $\Eph^\perp\propto\kperp^{-3}$. One can obtain such a spectrum if one 
assumes that the fluctuation energy present {\em at each scale}, not just at 
the outer scale, is determined by the balance between the instability growth rate, 
the nonlinear decorrelation rate---and also the phase mixing, which removes 
the energy to high $m$'s, so there is no need for a constant flux.
Then each scale behaves as the outer scale described in \secref{sec:outer} 
($\ephi\propto\kperp^{-1}$, as in \eqref{eq:phio}). We consider this scenario 
much too fanciful (it would require quite a complicated set of arrangements 
in the $(\kperp,\kpar)$ space) and rather unlikely for a system 
far from the threshold. Note also that the restriction $\xi<1$ would not 
apply in a system where the energy injection rate is not proportional to 
$\kperp$, e.g., one where $\chi$ in \eqref{eq:g} is just a large-scale force
and so the injection occurs only at the scale of the force. Then the non-universal 
spectrum \exref{eq:Ephi_xi} can be steeper than $\kperp^{-3}$, although 
we must have $\xi<4$ in order for $\kperp\uperp$ to increase with $\kperp$ 
and so for the nonlinear transfer to stay local. The steepest possible 
spectrum in this case is, therefore, $\Eph^\perp\propto\kperp^{-5}$.} 
However, numerical---or, indeed, experimental---evidence does not appear to  
support spectra that are significantly steeper 
than $\kperp^{-2}$ at long (above the Larmor scale) wavelengths 
\citep[e.g.,][]{hennequin04,goerler08,casati09,vermare11,barnes11,kobayashi15}. 
Furthermore, numerical investigations by \citet{teaca12,teaca14} and \citet{banon14} 
confirm local nonlinear energy transfer and possibly even 
constant fluxes, albeit with a number of caveats regarding non-asymptoticity 
of the simulations, consequent possible non-universality of their results, 
as well as distinctly measurable, if not dominant, amounts of dissipation 
(meaning, in their context, phase mixing) everywhere.
 
In what follows, we shall see that, in a sufficiently collisionless plasma, 
the constant-flux assumption is safer than it might appear. 
 
\subsection{Hermite ``cascade''}
\label{sec:Hcascade}

As the last bit of essential background, let us consider what happens with free 
energy in phase space if we treat phase mixing as the dominant process and 
ignore nonlinearity---the opposite extreme to that pursued in \secref{sec:fluid_theory}. 

Returning to \eqref{eq:gm} and dropping the advection term $\vuperp\cdot\vdperp$ 
for the time being, we perform a Fourier transform in the parallel direction 
and introduce the following very useful functions \citep{zocco11}: 
\beq
\tg_m(\kpar) = (i\,\sgn\,\kpar)^m g_m(\kpar), 
\label{eq:tg_def}
\eeq
where $g_m(\kpar)$ are the Fourier--Hermite harmonics. 
The (linearised) \eqref{eq:gm} then becomes
\beq
\frac{\dd\tg_m}{\dd t} + 
\frac{|\kpar|\vth}{\sqrt{2}}(\sqrt{m+1}\,\tg_{m+1} - \sqrt{m}\,\tg_{m-1}) = -\nu m\tg_m. 
\label{eq:tg}
\eeq

The point of these manipulations is that they have made the 
phase-mixing term on the left-hand side of \eqref{eq:tg} look like 
a derivative with respect to $m$. Indeed, assuming, in the limit of $m\gg1$, 
that we can treat $\tg_m$ as though it were continuous and differentiable in $m$ 
(an assumption that will come under close scrutiny in \secref{sec:unmix}), i.e.,
$\tg_{m\pm 1}\approx \tg_m \pm \dd_m\tg_m$, we have
\begin{align}
\nonumber
\sqrt{m+1}\,\tg_{m+1} - \sqrt{m}\,\tg_{m-1} &= 
\sqrt{m}\lt(\sqrt{1+\frac{1}{m}}\,\tg_{m+1} - \tg_{m-1}\rt)\\
&\approx \sqrt{m}\lt(\frac{\tg_m}{2m} + 2\frac{\dd\tg_m}{\dd m}\rt)
= 2m^{1/4}\frac{\dd}{\dd m}m^{1/4}\tg_m. 
\end{align}
Thus, \eqref{eq:tg} becomes
\beq
\frac{\dd\tg_m}{\dd t} 
+ \sqrt{2}\,|\kpar|\vth\, m^{1/4}\frac{\dd}{\dd m}m^{1/4}\tg_m = - \nu m\tg_m.
\eeq
Introducing the Fourier-Hermite free-energy spectrum 
$C_m(\kpar)=\la|\tg_m(\kpar)|^2\ra = \la|g_m(\kpar)|^2\ra$, we find 
\beq
\frac{\dd C_m}{\dd t} 
+ \frac{\dd}{\dd m}|\kpar|\vth \sqrt{2m}\,C_m = - 2\nu m C_m.
\label{eq:Cm_naive}
\eeq
In steady state, the solution is (\citealt{zocco11}; cf.\ \citealt{watanabe04}) 
\beq
C_m = \frac{A(\kpar)}{\sqrt{m}}\,e^{-(m/\mc)^{3/2}},\quad
\mc = \lt(\frac{3|\kpar|\vth}{2\sqrt{2}\,\nu}\rt)^{2/3}, 
\label{eq:Cm_sln}
\eeq
where $A(\kpar)$ is the constant of integration. 
Below the collisional cutoff, $m\ll\mc$, the power-law scaling $C_m\propto m^{-1/2}$
is the solution corresponding to constant free-energy flux in Hermite space 
(the Hermite flux is the expression under $\dd_m$ in \eqref{eq:Cm_naive}). 
It is possible to show quite rigorously (by direct Hermite transformation 
of the Landau response function) that this is indeed the Hermite-space 
solution that arises in a linear system with external forcing at low $m$ 
and Landau damping \citep{kanekar15}. 

The solution \exref{eq:Cm_sln} has two important properties.  
Firstly, the free-energy dissipation associated with it (the last term in \eqref{eq:Wkin})
is dominated by Hermite moments with $m\sim\mc$ 
and does not explicitly depend on the collision frequency (assuming $A(\kpar)$ does not), 
\beq
D = \nu\sum_m m\la g_m^2\ra = 
\nu\sum_{\kpar}\sum_m mC_m(\kpar) 
\approx \nu\sum_{\kpar}\int_{\sim 1}^\infty\rmd m\, mC_m(\kpar) = 
\sum_{\kpar} \frac{|\kpar|\vth A(\kpar)}{\sqrt{2}}. 
\label{eq:diss_lin}
\eeq
Secondly, the total amount of free energy stored in the phase space in order 
to achieve this finite dissipation (corresponding to finite amount of injected power) 
diverges as $\nu\to 0$: 
\beq
W \approx \frac{1}{2}\sum_{\kpar}\int_{\sim 1}^\infty\rmd m\, C_m(\kpar) 
= \sum_{\kpar}\frac{\Gamma(1/3)}{3^{2/3}\sqrt{2}}\frac{A(\kpar)}{(|\kpar|\vth)^{2/3}\nu^{1/3}} 
\label{eq:Wlin}
\eeq
\citep{kanekar15}. 

Thus, if we thought that Landau damping in a turbulent system works in the same 
way as it does in a linear one, we might have to conclude that,  
rather than staying in low $m$'s and being nonlinearly cascaded to small spatial 
scales, as in a fluid problem, the free energy fills up phase space and dissipates 
on collisions. 
A dedicated study of the Hermite spectra of slab ITG turbulence by \citet{hatch13,hatch14} 
showed that this does {\em not} happen, with Hermite spectrum of the free energy 
following a much steeper power law than \eqref{eq:Cm_sln} and the wavenumber spectrum 
consistent with \eqref{eq:Ephi1D}. 
In what follows, we will show how such a solution can emerge 
(\secref{sec:1D} has the answer and \apref{ap:threshold} 
the physical basis for it; see \secref{sec:totals} for the nonlinear 
versions of \eqsand{eq:Wlin}{eq:diss_lin}).  

\section{Formalism}
\label{sec:formalism}

\subsection{Phase mixing and anti-phase-mixing}
\label{sec:unmix}

Our first order of business in constructing an appropriate mathematical description 
for phase-space turbulence is to reexamine our rather blithe assumption in \secref{sec:Hcascade} 
that $\tg_m(\kpar)$, defined by \eqref{eq:tg_def} and satisfying \eqref{eq:tg} 
(to which the nonlinearity will be restored in \secref{sec:nlin}), 
can be treated as continuous in~$m$. 

Consider 
\beq
1\ll m \ll \lt(\frac{|\kpar|\vth}{\nu}\rt)^2. 
\label{eq:m_cont}
\eeq
If we assume that the rate 
of change of $\tg_m$ is small compared to $\sqrt{m}\,|\kpar|\vth$, 
\eqref{eq:tg} tells us that, to lowest approximation,  
\beq
\sqrt{m+1}\,\tg_{m+1} - \sqrt{m}\,\tg_{m-1} = 0
\quad\Rightarrow\quad
\tg_{m+1}\approx\tg_{m-1}. 
\eeq
This has two solutions: 
\beq
\tg_{m+1}\approx\pm\tg_m,
\label{eq:gm_alt}
\eeq
so, in fact, either $\tg_m$ or $(-1)^m\tg_m$ can be treated as continuous in $m$. 
We therefore introduce the following decomposition (which we already used in \citealt{kanekar15} 
and \citealt{parker15})
\beq
\tg_m = \tg_m^+ + (-1)^m\tg_m^-,
\label{eq:decomp}
\eeq
where 
\beq
\tg_m^+ = \frac{\tg_m + \tg_{m+1}}{2}, 
\quad
\tg_m^- = (-1)^m\frac{\tg_m-\tg_{m+1}}{2}
\label{eq:gpm_def}
\eeq
can both be assumed continuous in~$m$. Evolution equations for these two types 
of modes can be derived by adding or subtracting evolution equations \exref{eq:tg} 
for $\tg_m$ and $\tg_{m+1}$ and then expanding in large $m$ in the same fashion 
as we did in \secref{sec:Hcascade}. The result~is
\beq
\frac{\dd\tg_m^\pm}{\dd t} 
\pm \sqrt{2}\,|\kpar|\vth\, m^{1/4}\frac{\dd}{\dd m}m^{1/4}\tg_m^\pm = - \nu m\tg_m^\pm.
\label{eq:gpm_lin}
\eeq
Manifestly, the ``$+$'' modes are the phase-mixing modes, propagating from small to large $m$, 
whereas the ``$-$'' modes propagate from large to small $m$ and thus represent 
``anti-phase-mixing'': free energy coming back from phase space, a possibility 
earlier mooted, in somewhat different terms, by \citet{hammett93} and \citet{smith97}. 
We shall discuss the energetics of this process more quantitatively 
in \secref{sec:energetics}

In a linear problem, in the absence of free-energy sources at high $m$, 
the only solution that satisfies the boundary condition $\tg_{m\to\infty}\to0$ is 
$\tg_m^-=0$, so there will be no anti-phase-mixing and the treatment in \secref{sec:Hcascade}
is correct.\footnote{\citet{kanekar15} showed that in a (forced) linear problem, 
the spectrum of the ``$-$'' modes is $\propto m^{-3/2}$ and so subdominant to the 
spectrum \exref{eq:Cm_sln} of the ``$+$'' modes. This does not mean that there is 
some small subdominant amount of anti-phase-mixing in a linear system, but is rather 
due to the interpretation of $\tg_m^+$ and $\tg_m^-$ as being forward and backward 
propagating modes in $m$ space being correct only to lowest order in~$1/m$. Note that 
this interpretation breaks down also at such large $m$ that the inequality 
\exref{eq:m_cont} is no longer satisfied. When $m\gg(|\kpar\vth|/\nu)^2$, 
the collisional term in the right-hand side of \eqref{eq:tg} is dominant
and the solution is $\tg_m\approx(|\kpar|\vth/\nu\sqrt{2m})\tg_{m-1}\ll\tg_{m-1}$. 
Therefore, in this approximation, the two modes are $\tg_m^+\approx\tg_m/2\approx(-1)^m\tg_m^-$, 
and so they formally have the same energy.} 
As we are about to see, the situation changes once nonlinearity is accounted~for. 

\subsection{Nonlinear coupling and plasma echo}
\label{sec:nlin}

Let us now restore the nonlinear advection (the second term on the left-hand side 
of \eqref{eq:gm}) and Fourier transform it in the parallel direction: 
\beq
\lt(\frac{\dd g_m}{\dd t}\rt)_\mathrm{nl}\!\!\!
= -\lt[\vuperp\cdot\vdperp g_m\rt](\kpar) 
= -\!\!\!\!\!\!\!\sum_{\ppar+\qpar=\kpar}\!\!\!\!\! 
\vuperp(\ppar)\cdot\vdperp g_m(\qpar). 
\eeq
Then the nonlinear term that must be added to the right-hand side of 
\eqref{eq:tg}~is 
\beq
\lt(\frac{\dd \tg_m}{\dd t}\rt)_\mathrm{nl}\!\!\!
= -(i\,\sgn\,\kpar)^m\lt[\vuperp\cdot\vdperp g_m\rt](\kpar) 
= -\!\!\!\!\!\!\!\sum_{\ppar+\qpar=\kpar}\!\!\!\!\! 
\vuperp(\ppar)\cdot\vdperp 
\lt(\frac{\sgn\,\kpar}{\sgn\,\qpar}\rt)^m \tg_m(\qpar). 
\label{eq:tg_nlin}
\eeq
Finally, adding or subtracting the above for the $m$-th and $(m+1)$-st 
Hermite moments, and using the decomposition \exref{eq:decomp}, 
we find the nonlinear term for \eqref{eq:gpm_lin}: 
\begin{align}
\nonumber
\frac{\dd\tg_m^\pm}{\dd t} 
\pm \sqrt{2}\,|\kpar|\vth\, & m^{1/4}\frac{\dd}{\dd m}m^{1/4}\tg_m^\pm 
+ \nu m\tg_m^\pm =\\ 
&-\!\!\!\!\!\!\!\sum_{\ppar+\qpar=\kpar}\!\!\!\!\! 
\vuperp(\ppar)\cdot\vdperp 
\lt[\delta^+_{\kpar,\qpar}\tg_m^\pm(\qpar) 
+ \delta^-_{\kpar,\qpar}\tg_m^\mp(\qpar)\rt],
\label{eq:gpm}
\end{align}
where $\delta^{\pm}_{\kpar,\qpar}=\lt[1\pm\sgn(\kpar\qpar)\rt]/2$, 
i.e., $\delta^+$ is non-zero (and equals unity) only if $\kpar$ and $\qpar$ have the same 
sign, and $\delta^-$ is non-zero (and equals unity) only if they have the opposite sign. 

The key development manifest in \eqref{eq:gpm}
is that the advecting velocity field can couple parallel 
wave numbers of opposite signs and thus produce anti-phase-mixing ``$-$'' modes 
out of phase-mixing ``$+$'' ones and vice versa; 
$\tg_m^-=0$ is no longer a valid solution. 
This is a manifestation of the textbook plasma-physics phenomenon known as 
plasma echo \citep{gould67,malmberg68}. 
The importance of it in our discussion is that once the free-energy flux 
through phase space is not compelled to be unidirectional towards high $m$'s 
(as it was in the naive treatment of \secref{sec:Hcascade}), all bets 
are off as to the effectiveness of Landau damping/phase mixing as 
a dissipation mechanism in a nonlinear system.  

\subsection{Dual kinetic equation in phase space}
\label{sec:dual}

\Eqref{eq:gpm} can be recast in a remarkably simple form if we introduce 
a change of variables and a rescaling of $\tg_m^\pm$: 
\beq
s = \sqrt{m},
\quad
\tf(s,\kpar) = m^{1/4}\cdot\lt\{\begin{array}{lcl} 
\tg_m^+(\kpar) &\mathrm{if}& \kpar\ge 0,\\
\tg_m^-(\kpar) &\mathrm{if}& \kpar< 0.\\
\end{array}\rt.
\label{eq:tf_def}
\eeq
For any given $\kpar\ge 0$, the original distribution 
function is reconstructed in the following way, via 
\eqsand{eq:tg_def}{eq:decomp}:  
\beq
g_m(\kpar) = (-i)^m\lt[\tf(\sqrt{m},\kpar) + (-1)^m\tf(\sqrt{m},-\kpar)\rt],
\quad g_m(-\kpar) = g_m^*(\kpar). 
\eeq
The new function $\tf$ satisfies
\beq
\frac{\dd\tf}{\dd t} + \frac{\kpar\vth}{\sqrt{2}}\frac{\dd\tf}{\dd s} + \nu s^2\tf
= -\sum_{\ppar} \vuperp(\ppar)\cdot\vdperp\tf(\kpar-\ppar). 
\label{eq:tf}
\eeq
The echo effect in this equation looks explicitly like mode coupling 
from positive to negative parallel wave numbers, or vice versa, whereas the 
phase mixing and anti-phase-mixing are simply propagation in $s$ 
with velocity $\kpar\vth/\sqrt{2}$. We will make repeated references 
to this equation in the scaling arguments of \secref{sec:scaling}. 

\Eqref{eq:tf} is a kinetic equation in phase space dual to the original kinetic 
equation \exref{eq:g}, with the variable $s$ (or $\sqrt{2}\,s/\vth$) effectively 
acting as a Fourier dual to $\vpar$---this is not a huge surprise because 
for~$m\gg1$, Hermite polynomials are well approximated by trigonometric functions 
in $\vpar$, with ``frequency'' $\sqrt{2m}/\vth$:  
\beq
H_m(\hvpar) e^{-\hvpar^2/2} \approx \sqrt{2}\lt(\frac{2m}{e}\rt)^{m/2}\!\!\!
\cos\lt(\frac{\sqrt{2m}}{\vth}\,\vpar - \frac{\pi m}{2}\rt). 
\eeq 

It is worth stressing that, while the functions $\tg_m^\pm(\kpar)$ are subject 
to reality conditions, inherited from $g_m$ via $\tg_m$ (see definitions 
\exref{eq:tg_def} and \exref{eq:gpm_def}), 
\beq
g_m(-\kpar) = g_m^*(\kpar)
\quad\Rightarrow\quad
\tg_m(-\kpar) = \tg_m^*(\kpar)
\quad\Rightarrow\quad
\tg_m^\pm(-\kpar) = \lt[\tg_m^\pm(\kpar)\rt]^*,
\label{eq:reality}
\eeq
the function $\tf(\kpar)$ has no such property because it has been spliced together
from the positive-$\kpar$ values of $\tg_m^+$ and the negative-$\kpar$ values 
of $\tg_m^-$ and there is, {\em a priori}, no symmetry between the ``$+$'' 
and ``$-$'' modes. 

Let us reinforce this point by showing that a solution of 
\eqref{eq:tf} can only have the property 
\beq
\tf(-\kpar)=\tf^*(\kpar)
\label{eq:tf_reality}
\eeq
if the phase mixing is ignorable (this is worth noting 
because if \eqref{eq:tf_reality} does hold, then the free-energy flux 
in Hermite space vanishes, as per \eqref{eq:Gamma_F}; we will make good 
use of this argument in \secref{sec:fluid_region}).  
Taking the complex conjugate of \eqref{eq:tf} and subtracting from it 
the same equation written for $\tf(-\kpar)$, we get
\begin{align}
\nonumber
\lt(\frac{\dd}{\dd t} + \nu s^2\rt)&\lt[\tf^*(\kpar)-\tf(-\kpar)\rt]
+\frac{\kpar\vth}{\sqrt{2}}\frac{\dd}{\dd s}\lt[\tf^*(\kpar)+\tf(-\kpar)\rt] =\\ 
&\qquad\qquad-\sum_{\ppar}\vuperp(\ppar)\cdot\vdperp\lt[\tf^*(\kpar+\ppar) - \tf(-\kpar-\ppar)\rt], 
\label{eq:not_real}
\end{align} 
where have used $\vuperp^*(\ppar)=\vuperp(-\ppar)$ and then changed the summation 
variable $\ppar\to-\ppar$ in the sum involving $\tf^*$.  
\Eqref{eq:not_real} is compatible with the condition \exref{eq:tf_reality} 
only if the phase mixing term can be ignored---which might happen because 
$\kpar$ is small and/or because $\tf$ depends on $s$ in such a way that 
the phase-mixing term is subdominant at, say, high~$s$. 

\subsection{Free-energy spectrum and free-energy flux}
\label{sec:energetics}

Since we are going to discuss free-energy spectra and free-energy fluxes 
in phase space, let us provide the formal definitions and evolution equations for them. 

We define, in the same way as we did in \secref{sec:Hcascade}, 
\beq
C_m(\kpar) = \la|\tg_m(\kpar)|^2\ra = \la|g_m(\kpar)|^2\ra. 
\eeq
Then, using \eqref{eq:tg} with the nonlinear term given by 
\eqref{eq:tg_nlin}, we have
\beq
\frac{\dd C_m}{\dd t} + \Gamma_m - \Gamma_{m-1} + 2\nu m C_m 
= 2\Re\lt\la\lt(\frac{\dd \tg_m}{\dd t}\rt)_\mathrm{nl}\tg_m^*\rt\ra
\equiv\lt(\frac{\dd C_m}{\dd t}\rt)_\mathrm{nl},
\label{eq:Cm_exact}
\eeq
where the free-energy flux in Hermite space is \citep[cf.][]{watanabe04}
\beq
\Gamma_m(\kpar) = \sqrt{2(m+1)}\,|\kpar|\vth \Re\la\tg_{m+1}\tg_m^*\ra
= \sqrt{2(m+1)}\,\kpar\vth \Im\la g_{m+1} g_m^*\ra. 
\label{eq:Gamma_def}
\eeq
These expressions are exact. In the limit of $m\gg1$, both $C_m$ and 
$\Gamma_m$ are continuous in $m$ (even if $\tg_m$ alternates sign, 
\eqref{eq:gm_alt}), so we may rewrite \eqref{eq:Cm_exact} as follows  
\beq
\frac{\dd C_m}{\dd t} + \frac{\dd\Gamma_{m}}{\dd m} + 2\nu m C_m 
=\lt(\frac{\dd C_m}{\dd t}\rt)_\mathrm{nl}.
\label{eq:Cm_cont}
\eeq
Using the definition \exref{eq:gpm_def} of the ``$\pm$'' modes
and defining their spectra
\beq
C_m^\pm(\kpar) = \la|\tg_m^\pm(\kpar)|^2\ra, 
\eeq 
we notice that, still exactly, for any $m$, 
\beq
\Gamma_m = \sqrt{2(m+1)}\,|\kpar|\vth\lt(C_m^+ - C_m^-\rt).
\label{eq:Gamma}
\eeq
Thus, the Hermite flux is {\em exactly} proportional to the difference between 
the spectra of the ``$+$'' and ``$-$'' modes. 
The sum of these spectra is the free energy, but only approximately, for $m\gg1$: 
\beq
C_m^+ + C_m^- = \frac{C_m + C_{m+1}}{2} \approx C_m. 
\eeq


The evolution equations for $C_m^\pm$ at large $m$ are best obtained from 
\eqref{eq:gpm}: 
\begin{align}
\nonumber
\frac{\dd C_m^\pm}{\dd t} \pm &|\kpar|\vth\,\frac{\dd}{\dd m}\sqrt{2m}\,C_m^\pm 
+ 2\nu m C_m^\pm =\\ 
&-2\Re\!\!\!\!\!\!\!\sum_{\ppar+\qpar=\kpar}\!\!\!\!\! 
\lt\la\lt[\tg_m^\pm(\kpar)\rt]^*\vuperp(\ppar)\cdot\vdperp 
\lt[\delta^+_{\kpar,\qpar}\tg_m^\pm(\qpar) 
+ \delta^-_{\kpar,\qpar}\tg_m^\mp(\qpar)\rt]\rt\ra.
\label{eq:Cpm}
\end{align}
The sum of these two equations gives us back \eqref{eq:Cm_cont} with 
$\Gamma_m$ given by \eqref{eq:Gamma} (with $m\gg1$). 
Another, more compact, way to write \eqref{eq:Cpm} is in terms of the spectrum of the 
function $\tf$ introduced in \secref{sec:dual}. Defining
\beq
F(s,\kpar) = \la|\tf(s,\kpar)|^2\ra,
\eeq
we infer from \eqref{eq:tf}:
\beq
\frac{\dd F}{\dd t} + \frac{\kpar\vth}{\sqrt{2}}\frac{\dd F}{\dd s} + 2\nu s^2 F 
= -2\Re \sum_{\ppar} \lt\la \tf^*(\kpar)\vuperp(\ppar)\cdot\vdperp\tf(\kpar-\ppar)\rt\ra.
\label{eq:F}
\eeq
Note that, whereas $C_m(\kpar)$, $C_m^\pm(\kpar)$ and $\Gamma_m(\kpar)$ must all be even in $\kpar$ 
because of the reality conditions~\exref{eq:reality}, there is no such constraint on 
$F(\kpar)$ and, in fact, it is the odd part of $F(\kpar)$ that sets the Hermite flux: 
in view of \eqref{eq:Gamma}, 
\beq
\Gamma_m(\kpar) \approx \sqrt{2}\,|\kpar|\vth\lt[F(s,|\kpar|) - F(s,-|\kpar|)\rt]. 
\label{eq:Gamma_F}
\eeq 

The next step in the formal solution of the problem is to solve 
\eqref{eq:F} for $F(s,\kpar)$. However, even in principle, this is only possible 
if a suitable closure is found for the triple correlator in the 
right-hand side. A particular solvable model will be discussed in \citet{scalar},
but it will come at the price of decoupling the advecting velocity from the 
advected distribution function (i.e., considering a ``kinetic passive scalar'', 
rather than the fully self-consistent turbulence problem). 
In general, as always with turbulence problems, we are reduced to (or blessed with) 
having to resort to phenomenological scaling theories, which we will pursue 
in the next section. 

\section{Scaling theory}
\label{sec:scaling}

In constructing the scaling theory for our turbulence in phase space, 
we shall continue to consider as sensible and valid 
the arguments in \secref{sec:outer} that led to 
estimates of the outer scale (\eqref{eq:kyo})
and the amplitude of $\ephi$ at that scale (\eqref{eq:phio}). 
We will, therefore, focus on what happens in the inertial range. 

The argument that is presented in this section is quite long because, 
even within the inertial range, the phase space splits into several regions, 
where different physics are at work---and building the full picture involves 
investigating each of these regions and matching free-energy spectra at their 
boundaries. A road map to what is done where is provided by the subsection 
headings and by the overall summary in \secref{sec:sum}, which an impatient 
reader might find it useful to read first. 

In what follows, wherever our expressions appear 
to be dimensionally incorrect, this is because the wave numbers are normalised 
to the outer scale: 
\beq
\frac{\kpar}{\kparo}=\kpar\Lpar \to \kpar,\quad
\frac{\kperp}{\kperpo} \to \kperp; 
\eeq
we will also omit, wherever this makes exposition more rather than less transparent, 
such dimensional factors as $\vth$, $\rho_i$, etc. 
We remind the reader that at the outer scale, the parallel-propagation/phase-mixing 
and the nonlinear-advection time scales are assumed comparable, 
$\kparo\vth\sim\kperpo\uperpo$. 

\subsection{Spectra in the phase-mixing-dominated region}
\label{sec:unfettered}

The first two terms on the left-hand side of \eqref{eq:tf} describe propagation of 
a perturbation in $s$ with time, along the characteristic
\beq
s = \frac{\kpar\vth}{\sqrt{2}}\,t. 
\eeq
If we consider $\kpar>0$, perturbations will phase-mix in an unfettered way 
for at least a time comparable to the time it takes the nonlinearity to couple 
these perturbations to different wave numbers:
\beq
t \lesssim \tauc \sim (\kperp\uperp)^{-1} \propto \kperp^{-r},  
\eeq  
where $r$ is the scaling exponent of the nonlinear decorrelation rate. 
This means that whatever spectrum, denoted $\Eph(\kpar,\kperp)$, 
prevails at low $s$ (and so low $m$), it will simply propagate to 
higher $s$ as long as 
\beq
s \lesssim \kpar\vth\tauc \sim \frac{\kpar\vth}{\kperp\uperp} \sim \frac{\kpar}{\kperp^r}.
\eeq
We can rearrange this statement to mean that, for any given $m=s^2$, 
the part of the wave-number space satisfying 
\beq
\kpar\vth\gtrsim\sqrt{m}\,\kperp\uperp
\quad\Leftrightarrow\quad
\kpar \gtrsim \sqrt{m}\,\kperp^r
\label{eq:CBm}
\eeq
will contain an exact replica of the low-$m$ spectrum:\footnote{We assume 
that there is no discontinuity in the Hermite spectrum at low $m$, i.e., that 
the low-$s$ limit of the solution to \eqref{eq:tf} (which is technically only 
valid for $s=\sqrt{m} \gg 1$) will smoothly connect onto the spectra of low-$m$ ``fluid'' 
moments $\ephi$, $\upar$, $\dTpar$, etc.\ and also that the spectra of these 
quantities all have the same scaling with $\kpar$ and $\kperp$.} 
\beq
\label{eq:replica}
E_{\tf}(s,|\kpar|,\kperp) = \sqrt{m}\,E^+_m(\kpar,\kperp) \sim 
\Eph(\kperp,\kpar) \sim \kperp^b\kpar^{-a}.
\eeq
Here we have defined 2D spectra
\begin{align}
\nonumber
E_{\tf}(s,\kpar,\kperp) &= 2\pi\kperp\la|\tf(s,\kpar,\vkperp)|^2\ra,\\
E^\pm_m(\kpar,\kperp) &= 2\pi\kperp\la|\tg^\pm_m(\kpar,\vkperp)|^2\ra,\\
E_\ephi(\kpar,\kperp) &= 2\pi\kperp\la|\ephi(\kpar,\vkperp)|^2\ra,
\nonumber
\end{align}
where $\ephi(\kpar,\vkperp)$ etc.\ are Fourier transforms of 
the original fields in all three spatial directions. 
We shall refer to the lower bound on $\kpar$ (or upper bound on $\kperp$) 
defined by the condition \exref{eq:CBm}, $\kpar\sim\sqrt{m}\,\kperp^r$,  
as the {\em phase-mixing threshold}. 

The scaling exponents $a$ and $b$ in \eqref{eq:replica} are as yet unknown. 
One of them, $b$, can be determined in a purely 
``kinematic'' way: since it describes the low-$\kperp$ (see \eqref{eq:CBm}) 
asymptotic behaviour of the spectrum, it must, in a homogeneous isotropic system, 
be $b=3$ (the derivation of this result, which is quite standard, 
is given in \apref{ap:k3}---it describes the spectrum at perpendicular 
wavelengths that are longer than the perpendicular correlation scale 
of perturbations with a given $\kpar$). 

Thus, we have found a {\em phase-mixing-dominated region} 
(as we shall henceforth call it) of the phase space, with spectra
\beq
E_m^+(\kpar,\kperp) \sim \frac{\kperp^3\kpar^{-a}}{\sqrt{m}},\quad
E_m^-(\kpar,\kperp) \ll E_m^+(\kpar,\kperp),\quad
\kpar \gtrsim\sqrt{m}\,\kperp^r. 
\label{eq:E3ar}
\eeq 
These and all subsequent spectra that will emerge are sketched in \figref{fig:Em}, 
which the reader is invited to consult for illustration (and preview) of the 
upcoming results, as they emerge. 

Unsurprisingly, in \eqref{eq:E3ar} 
we have a $1/\sqrt{m}$ Hermite spectrum---the standard 
linear result already derived in \secref{sec:Hcascade}.  
The anti-phase-mixing component of the free energy ($E_m^-$) must be small 
compared to the phase-mixing one here because 
this is the part of phase space where the nonlinearity has no time 
to exert any influence and so there will not be any echo effect. 

While we do not yet know the exponent $a$ (it will be deduced, in two 
different ways, in \secsand{sec:scaling_phi}{sec:int}), 
it is clear that $E_m^+(\kpar,\kperp)$
must decay sufficiently fast with $\kpar$ in order for the total free energy 
not to diverge at short parallel wave lengths. 
This tendency for the free-energy spectrum to decay sharply 
at parallel wave numbers bounded from below (or, equivalently, at perpendicular 
wave numbers bounded from above) by the phase-mixing threshold
$\kpar\vth\sim\kperp\uperp$ (one might also call this threshold the 
``phase-space critical balance'')
was recently reported by \citet{hatch13,hatch14} (cf.\ \citealt{watanabe04})
in their simulations of slab ITG turbulence 
(they, however, had a different explanation for it).

\subsection{Spectra of low moments}
\label{sec:scaling_phi}

As we explained in \secref{sec:unfettered}, the spectrum \exref{eq:E3ar} 
is inherited (propagated by phase mixing) from low $m$'s, so we must have 
\beq
\Eph(\kpar,\kperp) \sim \kperp^3\kpar^{-a},\quad 
\kpar\gtrsim\kperp^r.
\label{eq:Ephi3a}
\eeq
Thus, this is the 2D spectrum of the electrostatic turbulence 
on the short-parallel-wavelength side 
of the critical-balance condition \exref{eq:CB}. 

As we argued in \secref{sec:CB}, the critical balance is essentially 
a causality condition and so the spectrum at the long-parallel-wavelength side 
of the critical balance, $\kpar\lesssim\kperp^r$, 
must reflect the fact that the perturbations at these parallel scales are 
essentially uncorrelated. The spectrum of such uncorrelated perturbations is 
the spectrum of white noise, so 
\beq
\Eph(\kpar,\kperp) \sim \kpar^{0}\kperp^{-c},\quad 
\kpar\lesssim\kperp^r.
\label{eq:Ephi0c}
\eeq
Matching this with \eqref{eq:Ephi3a} along the curve $\kpar\sim\kperp^r$ gives 
\beq
a = \frac{3+c}{r}. 
\label{eq:a_cr}
\eeq 

If $a>1$, then $\kpar\sim\kperp^r$ is the energy-containing parallel scale 
for any given $\kperp$. The 1D perpendicular spectrum is, therefore,
\beq
\Eph^\perp(\kperp) = \int\rmd\kpar \,\Eph(\kpar,\kperp)
\sim \int_0^{\kperp^r}\rmd\kpar\,\kpar^0\kperp^{-c} 
\sim \kperp^{-(c-r)}.
\label{eq:Ephi_perp_cr}
\eeq
This immediately implies a consistency relation between $c$ and $r$:\footnote{We remind the 
reader that $\ephi$ here and in all similar calculations in this paper is {\em not} the Fourier transform 
of the potential, but rather its amplitude corresponding to the scale $\kperp^{-1}$ (this can 
be thought of, for example, as the typical magnitude of the potential's increment across 
a distance $\kperp^{-1}$). Its relationship to the Fourier transform $\ephi_{\vk}$ and to the 
1D spectrum $\Eph^\perp(\kperp)$ was given in \eqref{eq:Ephi1D}. This can be understood dimensionally 
or by noticing that the energy associated with a given scale $\kperp^{-1}$ is the integral over 
the energies contained in the wave number $\kperp$ and larger, 
$\ephi^2 \sim\int_{\kperp}^\infty\rmd\kperp'\Eph^\perp(\kperp')\sim \kperp\Eph^\perp(\kperp)$ 
(the latter relation holds as long as the 1D spectrum is steeper than $\kperp^{-1}$).}
\beq
\kperp^r\sim\kperp\uperp\sim\kperp^2\ephi\sim\kperp^2(\kperp \Eph^\perp)^{1/2} 
\quad\Rightarrow\quad
r = 5 - c. 
\label{eq:r_c}
\eeq
Finally, the 1D parallel spectrum for any given $\kpar$ is dominated by $\kperp\sim\kpar^{1/r}$: 
\beq
\Eph^\parallel(\kpar) = \int\rmd\kperp\,\Eph(\kpar,\kperp)
\sim  \int_{\kpar^{1/r}}^\infty\rmd\kperp\,\kpar^0\kperp^{-c}
\sim \kpar^{-(c-1)/r}. 
\label{eq:Ephi_par_cr}
\eeq

\begin{figure}
\begin{center}
\begin{tabular}{ccc}
\raisebox{-0.5\height}{
\includegraphics[width=0.45\textwidth]{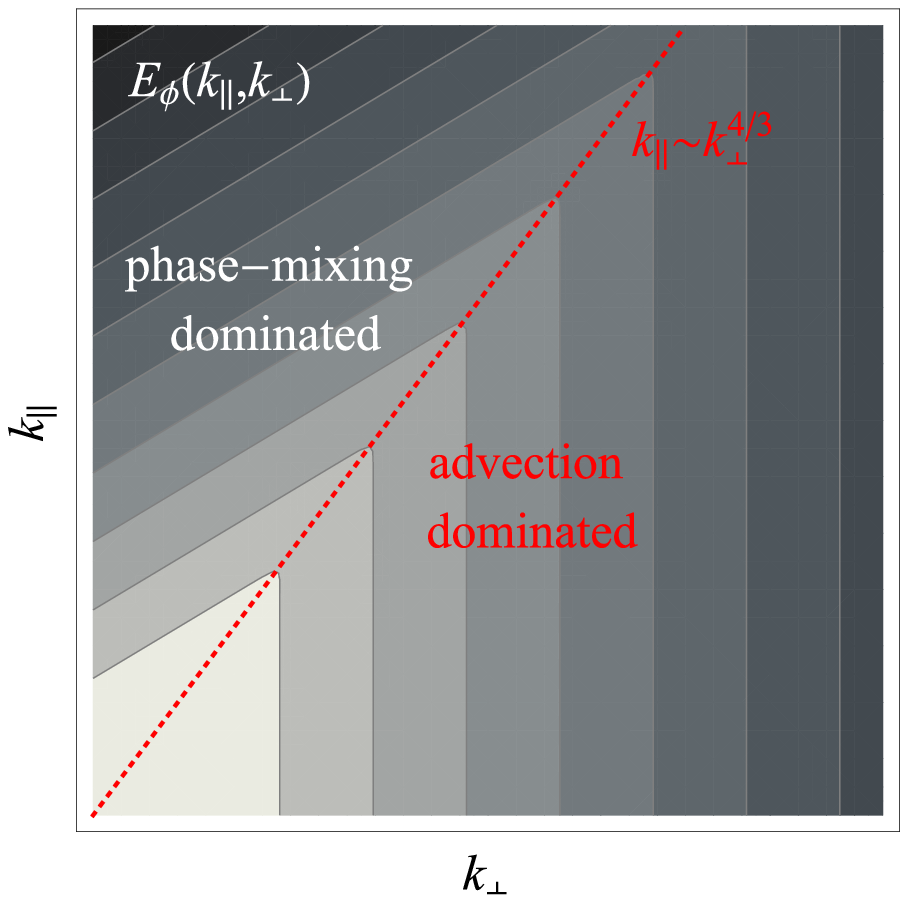}} &\qquad&
\begin{tabular}{c}
\includegraphics[width=0.45\textwidth]{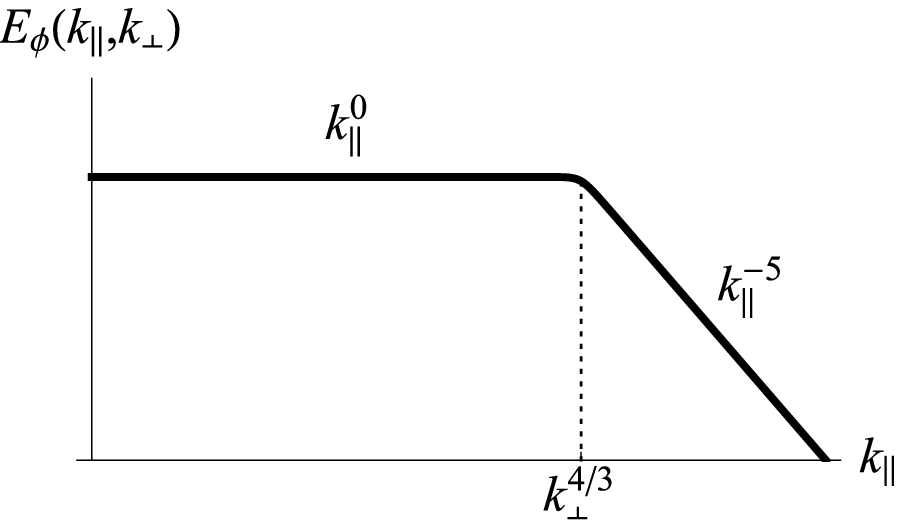}\\
(b)\\
\includegraphics[width=0.45\textwidth]{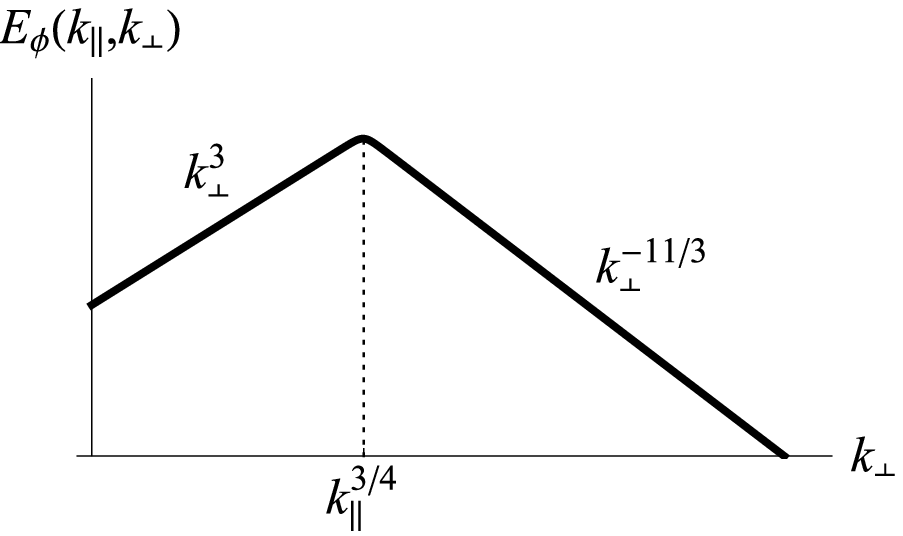}
\end{tabular}\\
(a) && (c)
\end{tabular}
\end{center}
\caption{Spectra of low Hermite moments, $\Eph(\kpar,\kperp)$: 
(a)~in the $(\kperp,\kpar)$ plane, 
(c)~vs.\ $\kpar$ at constant $\kperp$,
(b)~vs.\ $\kperp$ at constant $\kpar$.
All plots are logarithmic.}
\label{fig:Ephi}
\end{figure}

\subsubsection{Scaling exponents under constant-flux conjecture}

Note that so far, we have invoked no cascade physics, but, in order to determine 
the exponent $c$, we do now need to make an assumption as to how energy is 
passed from scale to scale. Energetically, only the wave-number region $\kpar\lesssim\kperp^r$ 
matters because at larger $\kpar$, the spectrum is assumed (and will be confirmed) 
to have a steep decay with $\kpar$ (\eqref{eq:Ephi3a}). 
We shall call it the {\em advection-dominated region} and anticipate that 
phase mixing there will not be a significant energy 
sink, i.e., the anti-phase-mixing energy flux due to the echo effect will 
on average cancel the phase-mixing flux, leading to effective conservation 
of $\la\ephi^2\ra$. Then we can return to the constant-flux argument of 
\secref{sec:in_range}: 
\beq
\ephi^2\kperp\uperp \sim \const
\quad\Rightarrow\quad
\kperp \Eph^\perp(\kperp) \sim \ephi^2 \sim \kperp^{-r}
\quad\Rightarrow\quad
c=1+2r,
\label{eq:c_r}
\eeq 
where \eqref{eq:Ephi_perp_cr} was used to obtain the last relation. 
Combining \eqsref{eq:c_r}, \exref{eq:r_c} and \exref{eq:a_cr}, we find 
\beq
r=\frac{4}{3},\quad
c=\frac{11}{3},\quad
a = 5. 
\label{eq:rca}
\eeq
This gives us back the \citet{barnes11} 1D spectra: 
\beq
\Eph^\perp(\kperp)\sim\kperp^{-7/3},\quad
\Eph^\parallel(\kpar)\sim\kpar^{-2}
\eeq
via \eqsand{eq:Ephi_perp_cr}{eq:Ephi_par_cr}, respectively.    
We have now also learned what the full 2D spectrum behind these 
1D ones is: combining \eqsand{eq:Ephi3a}{eq:Ephi0c} with 
the scaling exponents \exref{eq:rca}, we have
\beq
\Eph(\kpar,\kperp) \sim
\lt\{
\begin{array}{lcl}
\kpar^0\kperp^{-11/3} & \mathrm{if} & \kpar\lesssim\kperp^{4/3},\\
\kperp^3\kpar^{-5} & \mathrm{if} & \kpar\gtrsim\kperp^{4/3}. 
\end{array}
\rt.
\eeq 
These spectra are sketched in \figref{fig:Ephi}. 

Since we have fixed the value of the spectral exponent $a$ and since 
the spectra of $\ephi$ in the phase-mixing-dominated region, where 
this exponent applies, propagate to higher $m$'s, we now also have 
determined $E_m(\kpar,\kperp)$ for $\kpar\gtrsim\sqrt{m}\,\kperp^r$: 
see \eqref{eq:E3ar}. 

Physically, the validity of the argument that led to 
the last set of results (\eqref{eq:c_r} onwards) hinges on our ability to produce 
phase-space spectra that are consistent with substantial cancellation 
of the phase-mixing flux at $\kpar\lesssim\kperp^r$ and thus with the 
majority of the free energy residing in the low Hermite moments. 
Note that it is actually not controversial that the phase mixing 
should be negligible for $\kpar\ll\kperp^r$ because this means the 
phase-mixing rate is low compared to the nonlinear advection rate, 
$\kpar\vth\ll\kperp\uperp$, but, as we saw above, the energy 
is substantially dominated by the critical-balance curve $\kpar\sim\kperp^r$, 
where the two rates are comparable (recall our critique of the ``fluid'' theory 
in \secref{sec:inconsistent}). In what follows, we shall build a case for 
the spectra that we have just derived---and so we will carefully avoid 
using the constant-flux argument \exref{eq:c_r} and keep all the scaling 
exponents general.  

\subsection{Spectra of higher moments in the advection-dominated region}
\label{sec:fluid_region}

Let us now consider the higher Hermite moments, $m\gg1$.  
The condition for the phase-mixing rate 
to be negligible compared to the nonlinear advection rate is the opposite 
of the condition \exref{eq:CBm}: 
\beq
\kpar\vth\ll\sqrt{m}\,\kperp\uperp
\quad\Leftrightarrow\quad
\kpar \ll \sqrt{m}\,\kperp^r
\label{eq:no_phmix}
\eeq
(i.e., the part of the phase space on the side of the phase-mixing threshold 
opposite to the phase-mixing-dominated region). 
In \secref{sec:unfettered}, the phase-mixing threshold was derived by arguing 
that it represented the value of $s$ up to which the ``$+$'' perturbations at some given 
$\kpar$ and $\kperp$ could propagate before being diverted to different wave 
numbers by the nonlinear coupling. More generally and more formally, we may 
simply argue that, if, say, the $s$ dependence of $\tf$ is a power law 
(which it must be; see \apref{ap:threshold}), 
the size of the phase-mixing term in \eqref{eq:tf} can be estimated as $\sim(\kpar\vth/s)\tf$, 
which is negligible compared to the nonlinear term if the condition \exref{eq:no_phmix} 
is satisfied.

The advection-dominated region
\beq
\kpar\vth\lesssim\kperp\uperp
\quad\Leftrightarrow\quad
\kpar \lesssim \kperp^r,
\label{eq:fluid_region}
\eeq 
which, as we argued in \secref{sec:scaling_phi}, contains most of the energy content  
of the low-$m$ Hermite moments (collectively represented by $\ephi$) and, therefore, 
of the advecting velocity $\vuperp$, 
is well within the domain of validity of the condition \exref{eq:no_phmix}, provided $m\gg1$. 
Therefore, if we restrict our attention to the wave numbers \exref{eq:fluid_region}, 
we may neglect the phase-mixing term (second on the 
left-hand side) in \eqref{eq:tf} and thus deal with what is a purely ``fluid'' equation 
for~$\tf(s)$ at each $s$. As we argued at the end of \secref{sec:dual}, we can then have solutions 
satisfying $\tf^*(\kpar)=\tf(-\kpar)$, for which the ``$+$'' and ``$-$'' spectra are the 
same and the Hermite flux is zero (see \eqref{eq:Gamma_F}):  
\beq
E^+_m(\kpar,\kperp) \approx E^-_m(\kpar,\kperp). 
\eeq
Physically, this is because, 
in the advection-dominated region, the nonlinear coupling between 
positive and negative $\kpar$ mediated by the velocity field $\vuperp$ 
will be vigorous and fast---assuming, importantly, that interactions
between $\vuperp$ and $\tf$ are local in $\kpar$ and so, in the right-hand side of 
\eqref{eq:tf}, the sum $\sum_{\ppar}$ is dominated by wave-number triads with 
$\ppar\sim\kpar\sim\kpar-\ppar$.

With the phase mixing neglected, 
the variance of $\tf$ is (approximately) conserved at each $s$. 
The field $\tf(s)$ is nonlinearly cascaded to smaller scales (larger $\kperp$)
by the advecting velocity $\vuperp$, so the standard constant-flux argument gives us 
\beq
\tf^2(s)\kperp\uperp \sim \const(s) 
\quad\Rightarrow\quad
\tf^2(s) \propto \kperp^{-r}
\quad\Rightarrow\quad
E_m^\perp(\kperp) \propto \kperp^{-r-1}, 
\label{eq:Em_perp_r}
\eeq
where $E_m^\perp(\kperp)$ is the 1D perpendicular spectrum of $g_m$. 
Note that the spectrum has an $m$ dependence, which cannot be determined 
via this argument.

Since, in the advection-dominated region \exref{eq:fluid_region}, 
the parallel-communication times are long compared to the 
nonlinear-decorrelation times, 
the perturbations can be expected to have a white-noise spectrum 
in $\kpar$. Therefore, we can write their 2D spectrum as follows
\beq
E_m(\kpar,\kperp) \sim 
E_m^\pm(\kpar,\kperp) \sim \frac{\kpar^0\kperp^{-d}}{m^\sigma},
\quad \kpar \lesssim\kperp^r,  
\label{eq:Em_dsr}
\eeq   
where we have allowed for an as yet unknown scaling with $m$.
The $\kperp$-scaling exponent $d$ can be determined via the requirement that 
\eqref{eq:Em_dsr} be consistent with \eqref{eq:Em_perp_r}: 
assuming that, for any given $\kperp$, the region $\kpar\lesssim\kperp^r$ 
contains most of the energy,\footnote{Technically speaking, we do not 
yet know this. We will justify this assumption {\em a posteriori} in 
\secref{sec:int_spectra}.} 
\beq
E_m^\perp(\kperp) = \int\rmd\kpar\,E_m(\kpar,\kperp)
\sim \int_0^{\kperp^r}\rmd\kpar\,\frac{\kpar^0\kperp^{-d}}{m^\sigma}
\propto \kperp^{-(d-r)}
\quad\Rightarrow\quad
d = 1 + 2r.
\label{eq:d_r}
\eeq 

Since, as we have
argued here, the Hermite flux is (approximately) zero in this region 
of wave-number space for all higher $m$'s, there can be  
very little net free-energy flow out of low $m$'s---and so, 
in retrospect, we were justified in assuming in \secref{sec:scaling_phi} 
that the energy of the low $m$'s was conserved and so a constant-flux 
argument \exref{eq:c_r} could be used to deduce the scaling of $\ephi$. 
This allows us to adopt the scaling exponents \exref{eq:rca} (which 
we have thus far avoided using), in particular, $r=4/3$, 
and so, using \eqref{eq:d_r}, the free-energy spectrum is 
(see \figref{fig:Em} for illustration)
\beq
E^\pm_m(\kpar,\kperp) \sim \frac{\kpar^0\kperp^{-11/3}}{m^\sigma},
\quad \kpar \lesssim\kperp^{4/3}.  
\label{eq:Em_s}
\eeq
Unsurprisingly, there is continuity between the scalings of $E_m^\pm$ and 
the scaling of $\Eph$. The scaling exponent $\sigma$ will be found 
in \secref{sec:int_spectra}. 

Before moving on to complete our scaling theory, 
we note that jumping to the result \exref{eq:Em_s} already in this section was 
borne of pure impatience: we will discover in \secref{sec:int} that, in fact, 
it is possible to determine the scaling exponent $d$ without relying on
the as yet perhaps somewhat unconvincing claim that a constant-flux 
argument is legitimate for $\ephi$ despite the phase mixing being notionally 
not small along the critical-balance curve~$\kpar\sim\kperp^r$. 

\subsection{Intermediate region and matching conditions}
\label{sec:int}

We now have the form of the free-energy spectra in two regions, 
$\kpar\gtrsim\sqrt{m}\,\kperp^r$ (phase-mixing dominated, very little free energy, 
\eqref{eq:E3ar}) and $\kpar\lesssim\kperp^r$ 
(advection-dominated cascade, contains most of the free energy, \eqref{eq:Em_s}). 
It remains to determine the free-energy spectrum in the 
{\em intermediate region} between these two: 
\beq
E^+_m(\kpar,\kperp) \sim \frac{\kpar^{-a'}\kperp^{-d'}}{m^{\sigma'}},
\quad
\kperp^r\lesssim\kpar\lesssim\sqrt{m}\,\kperp^r
\label{eq:E_adsr}
\eeq 
(the following argument will only apply to the ``$+$'' modes; 
both the reasons for this and the way to determine the spectrum of the ``$-$'' 
modes will be explained in \secref{sec:minus}). 

We have three new scaling exponents, but we also have the requirement to match 
\eqref{eq:E_adsr} with \eqsand{eq:E3ar}{eq:Em_dsr} along the boundaries 
of the intermediate region. This gives us four relations
\beq
\sigma'=\sigma,\quad
d' + a'r = d,\quad
d' + a'r = ar-3,\quad
a' + 2\sigma' = 1+a,
\eeq
which we rearrange so:  
\beq
a' = \frac{d-d'}{r},\quad
d = ar-3,\quad 
\sigma = \sigma' = \frac{3 + r + d'}{2r}.
\label{eq:apds}
\eeq
Let us combine these with two equally uncontroversial 
(i.e., requiring no leaps of physical intuition)
matching and consistency relations 
from \secref{sec:scaling_phi}: \eqsand{eq:a_cr}{eq:r_c}, which we can 
rewrite as 
\beq
a = \frac{8-r}{r},\quad
c = 5-r.
\label{eq:ac}
\eeq 
The second of \eqsref{eq:apds} then gives 
\beq
d = 5 - r = c.
\eeq
Thus, the perpendicular scalings of 
$E_m^\pm$ and $\Eph$ must be the same in the advection-dominated region 
$\kpar\lesssim\kperp^r$. 
If we bring in \eqref{eq:d_r}, i.e., the constant-flux argument 
\exref{eq:Em_perp_r}, we get immediately 
\beq
r = \frac{4}{3},\quad
d = c = \frac{11}{3},\quad
a = 5.
\label{eq:rdca_final}
\eeq
These are the same as the exponents \exref{eq:rca}, except the need 
for the ``fluid'' constant-flux argument \exref{eq:c_r} for $\ephi$ has now been obviated
by the combination of a more solid phase-space argument \exref{eq:Em_perp_r}
and a number of inevitable consistency relations. 
\Eqsref{eq:apds} will give us $a'$, $\sigma'$ and $\sigma$ if we know $d'$. 
In order to determine the latter, we must now consider why physics 
in the intermediate phase-space region $\kperp^r\lesssim\kpar\lesssim\sqrt{m}\,\kperp^r$
should be at all different from what happens in 
the advection-dominated region $\kpar\lesssim\kperp^r$ considered in \secref{sec:fluid_region}
(and so why $d'\neq d$). 

\subsubsection{Spectra in the intermediate region}
\label{sec:int_spectra}

\begin{figure}
\begin{center}
\includegraphics[width=0.85\textwidth]{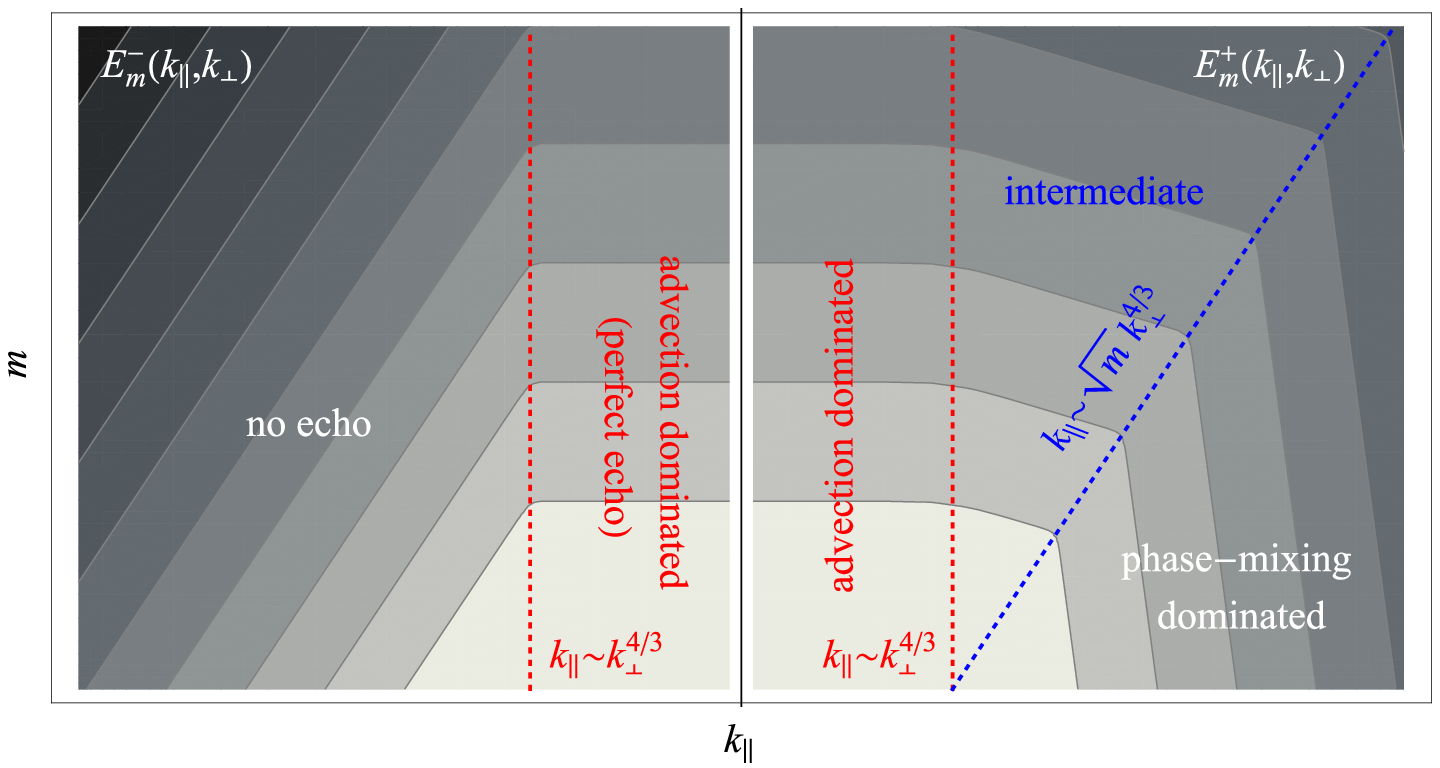}\qquad\qquad\quad\null\\
(a)
\begin{tabular}{ccc}
\raisebox{-0.5\height}{
\includegraphics[width=0.45\textwidth]{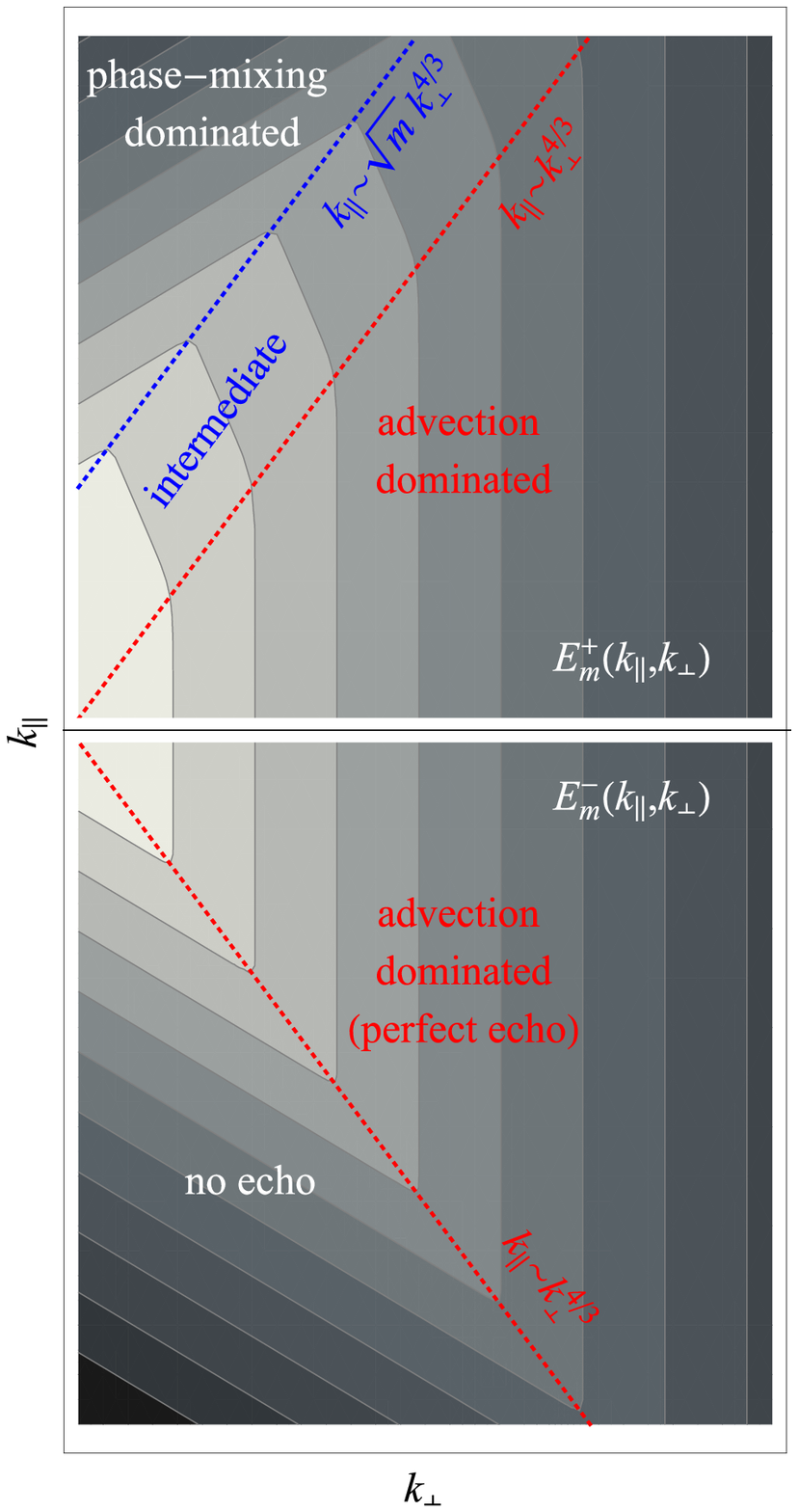}} &\qquad\qquad&
\begin{tabular}{c}
\includegraphics[width=0.45\textwidth]{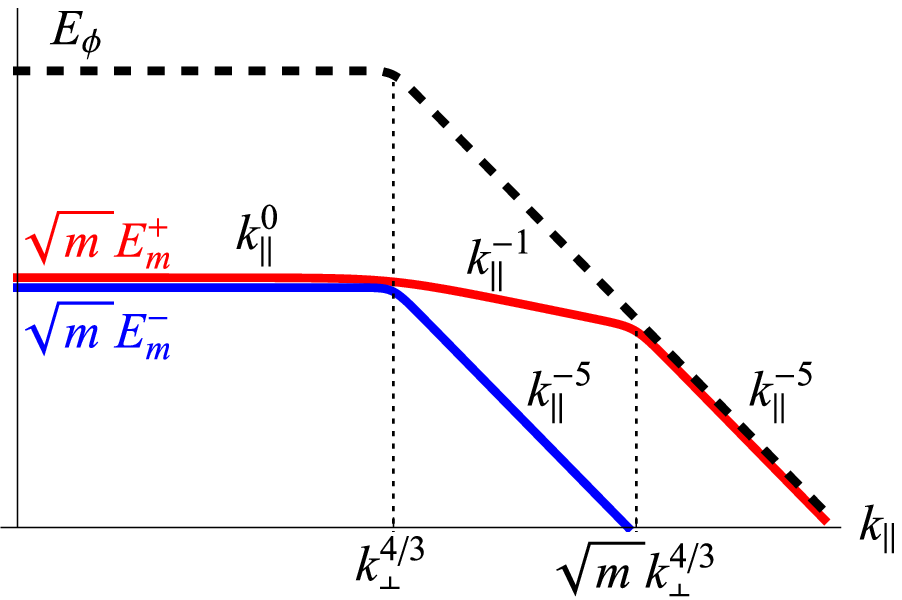}\\
(c)\\
\includegraphics[width=0.45\textwidth]{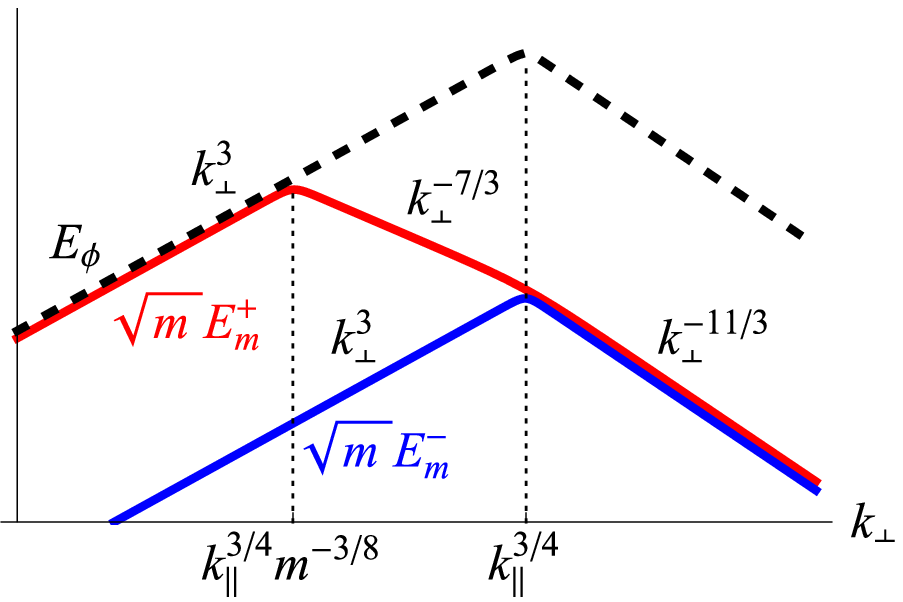}\\
(d)\\
\includegraphics[width=0.45\textwidth]{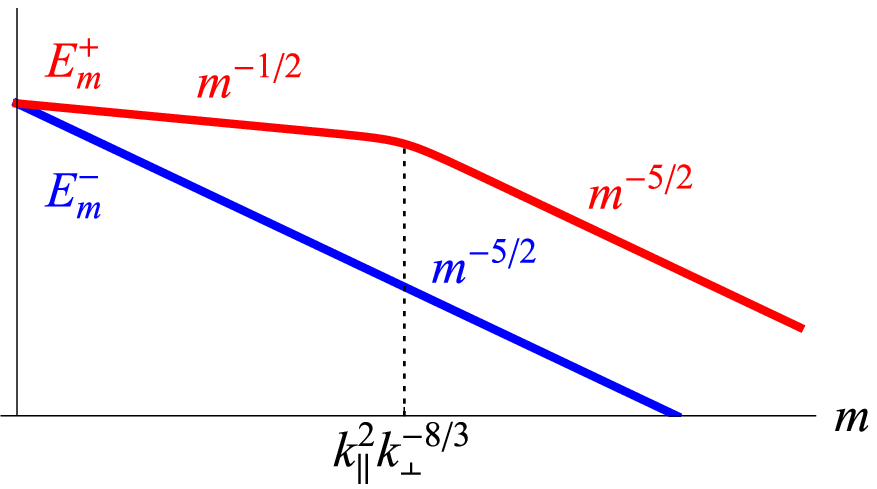}
\end{tabular}\\
(b) && (e)\\\\
\end{tabular}
\end{center}
\caption{Spectra of higher Hermite moments, $E_m^\pm(\kpar,\kperp)$: 
(a)~in the $(\kpar,m)$ plane at constant $\kperp$  
($E_m^+$ and $E_m^-$ are shown in the right and left panels, respectively; 
in the left panel, $\kpar$ increases leftwards),
(b)~in the $(\kperp,\kpar)$ plane at constant $m$ 
($E_m^+$ and $E_m^-$ are shown in the upper and lower panels, respectively; 
in the lower panel, $\kpar$ increases downwards), 
(c)~vs.\ $\kpar$ at constant $\kperp$ and $m$,
(d)~vs.\ $\kperp$ at constant $\kpar$ and $m$, 
(e)~vs.\ $m$ at constant $\kpar$ and $\kperp$ (such that 
$\kpar>\kperp^{4/3}$, otherwise the spectra are 
$E^\pm_m\sim m^{-5/2}$ at all $m$).
All plots are logarithmic. The spectrum $\Eph(\kpar,\kperp)$ 
(see \figref{fig:Ephi}) is given in (c) and (d) as a dashed line, 
for reference. Free-energy flow through phase space 
as represented in (a) and (b) is described in \secref{sec:flow_path}.}
\label{fig:Em}
\end{figure}

Except at the phase-mixing threshold $\kpar\sim\sqrt{m}\,\kperp^r$ 
(which is considered with more care in \apref{ap:threshold}), 
phase mixing in the intermediate region is dominated by nonlinear advection. However, 
interactions between $\vuperp$ and $\tf$ cannot, unlike in the advection-dominated 
region discussed in \secref{sec:fluid_region}, be local in $\kpar$. 
Indeed, we know from \secref{sec:scaling_phi} that there is very little 
energy left in $\vuperp$ at $\kpar\gg\kperp^r$. 
Assuming interactions to be local in $\kperp$, the wave-number 
sum in the right-hand side of \eqref{eq:tf} will be dominated by 
$\ppar\lesssim\kperp^r$ (see \apref{ap:nonlocal} for a careful analysis 
of the possible nonlocal interactions in the intermediate region). 
If $\kpar\gg\kperp^r$, then $\ppar\ll\kpar$, 
$\tf(\kpar-\ppar)\approx\tf(\kpar)$, and so \eqref{eq:tf} now describes 
the advection of $\tf(s,\kpar)$ by an essentially two-dimensional 
velocity field (its parallel scale is much longer than that of $\tf$), 
with both $s$ and $\kpar$ appearing as implicit parameters. 
This means that the variance of $\tf$ will be conserved for each 
individual $s$ and $\kpar$\footnote{This, incidentally, addresses a possible 
objection to the arguments in \secref{sec:fluid_region} that might have been 
troubling a perceptive reader: were we really justified in assuming that 
the conserved variance of $\tf(s)$ could all be accounted for within 
the region $\kpar\lesssim\kperp^r$, leading to the constant-flux 
argument \exref{eq:Em_perp_r}? The answer is that the free energy outside that 
region is either subdominant (\secref{sec:unfettered}) or conserved separately 
(\secref{sec:int_spectra}).}---and this in turn, by yet another 
constant-flux-in-$\kperp$ argument, 
implies
\beq
\tf^2(s,\kpar)\kperp\uperp \sim \const(s,\kpar)
\quad\Rightarrow\quad
\tf^2(s,\kpar)\propto \kperp^{-r}
\quad\Rightarrow\quad
E_m(\kpar,\kperp) \propto \kperp^{-r-1}.
\label{eq:Em_r} 
\eeq
This scaling is of the 2D spectrum, not of the 1D 
perpendicular one, because $\kpar$ is a fixed parameter, 
rather than a variable over which there can be any nonlinear coupling. 
Comparing \eqsref{eq:Em_r} and \exref{eq:E_adsr}, we read off $d'$ and hence, 
with the aid of the first and third \eqsref{eq:apds}, 
complete the determination of all scaling exponents: 
\beq
d' = r+1 = \frac{7}{3},\quad
a'=1,\quad
\sigma'=\sigma = \frac{5}{2}.
\label{eq:das_final}
\eeq
Thus, the free-energy spectrum in the intermediate region~is 
\beq
E^+_m(\kpar,\kperp) \sim \frac{\kpar^{-1}\kperp^{-7/3}}{m^{5/2}},
\quad
\kperp^{4/3}\lesssim\kpar\lesssim\sqrt{m}\,\kperp^{4/3},
\label{eq:Eint}
\eeq
sketched in \figref{fig:Em}. 
The spectrum of the ``$-$'' modes will be discussed in \secref{sec:minus}.

\subsubsection{1D spectra}
\label{sec:1D}

In the run up to \eqref{eq:d_r}, we assumed that integrating the 2D 
spectrum $E_m(\kpar,\kperp)$ with respect to $\kpar$ over the advection-dominated 
region $\kpar\lesssim\kperp^r$ captures most of the free energy contained 
in any fixed $\kperp$. Now in possession of \eqref{eq:Eint}, we see that 
this is not entirely correct: in fact, using now both \eqsand{eq:Em_s}{eq:Eint}, 
we find the 1D perpendicular spectrum to be 
\begin{align}
\nonumber 
E^{\perp+}_m(\kperp) &= \int\rmd\kpar\,E^+_m(\kpar,\kperp)\\
&\sim \int_0^{\kperp^{4/3}}\rmd\kpar\frac{\kpar^0\kperp^{-11/3}}{m^{5/2}}
+ \int_{\kperp^{4/3}}^{\sqrt{m}\,\kperp^{4/3}}\rmd\kpar\frac{\kpar^{-1}\kperp^{-7/3}}{m^{5/2}}
\sim\frac{\kperp^{-7/3}}{m^{5/2}}\lt(1+\ln\sqrt{m}\rt). 
\label{eq:Eperpm}
\end{align}
So there is logarithmically more free energy in the intermediate region, 
but this does not affect the $\kperp$ scaling, which is what we were after
in \eqref{eq:d_r}, so the derivation in \secref{sec:fluid_region} survives.
 
For completeness, let us also calculate the 1D parallel spectrum. Integration 
over $\kperp$ is dominated by the wave numbers around the 
phase-mixing threshold $\kperp\sim(\kpar/\sqrt{m})^{3/4}$,~so
\begin{align}
\nonumber
E^{\parallel+}_m(\kpar) &= \int\rmd\kperp\,E^+_m(\kpar,\kperp)\\
&\sim \int_0^{(\kpar/\sqrt{m})^{3/4}}\rmd\kperp\frac{\kperp^3\kpar^{-5}}{\sqrt{m}}
+ \int_{(\kpar/\sqrt{m})^{3/4}}^{\kpar^{3/4}}\rmd\kperp\frac{\kpar^{-1}\kperp^{-7/3}}{m^{5/2}}
\sim\frac{\kpar^{-2}}{m^2}. 
\label{eq:Eparm}
\end{align}
Note that this $m^{-2}$ scaling appears to be in decent agreement with the Hermite-space 
spectra reported by \citet{hatch13,hatch14}.\\ 

A perceptive reader might be feeling a growing resentment over our use 
of the spectrum \exref{eq:Eint} at wave numbers around the phase-mixing 
threshold $\kpar\sim\sqrt{m}\,\kperp^{4/3}$, even though,  
technically speaking, we have only justified \eqref{eq:Eint} in the 
region $\kperp^{4/3}\lesssim\kpar\ll\sqrt{m}\,\kperp^{4/3}$, seeing that 
at $\kpar\sim\sqrt{m}\,\kperp^{4/3}$,  
phase mixing cannot be neglected compared to the nonlinear advection. 
In \apref{ap:threshold}, we show that it nevertheless makes sense simply 
to match the spectrum \exref{eq:E_adsr} to the phase-mixing-dominated 
spectrum \exref{eq:E3ar} along the phase-mixing threshold. 

\subsection{Anti-phase-mixing spectra}
\label{sec:minus}

The arguments about the intermediate-region spectra presented in \secref{sec:int} only apply 
to the spectrum of the ``$+$'' modes. Since the advection velocity is effectively two-dimensional 
in the intermediate region (see \secref{sec:int_spectra}), there is no coupling between different 
parallel wave numbers and so no echo effect. Thus, if, as we argued in \secref{sec:fluid_region},
$E^-_m\approx E^+_m\propto\kperp^{-d}$ in the advection-dominated region, 
$\kperp\gtrsim\kpar^{1/r}$, due to vigorous 
coupling between parallel wave numbers, and $E^-_m\ll E_m^+$ in the intermediate 
and phase-mixing-dominated region, $\kperp\ll\kpar^{1/r}$, 
due to absence of any such coupling, 
we must expect that the energy-containing wave numbers for the ``$-$" modes 
are ones along the critical-balance curve $\kperp\sim\kpar^{1/r}$. 

The $\kperp\to0$ asymptotic behaviour of the ``$-$'' spectrum must be the same 
as for any other field, $E_m^-\propto\kperp^3$, because the reasons for it 
are purely kinematic (\apref{ap:k3}). Thus, we posit 
\beq
E_m^-(\kpar,\kperp)\sim \frac{\kperp^3\kpar^{-a''}}{m^{\sigma''}},
\quad\kpar\gtrsim\kperp^r
\eeq
and impose matching conditions between this spectrum and \eqref{eq:Em_dsr} 
at $\kpar\sim\kperp^r$:
\beq
a'' = \frac{d+3}{r}=5,\quad \sigma''=\sigma=\frac{5}{2}.
\eeq 
This completes the determination of the phase-space spectra of the anti-phase-mixing 
component of the free energy:
\beq
\label{eq:Eminus}
E_m^-(\kpar,\kperp)\sim \frac{1}{m^{5/2}}\cdot
\lt\{\begin{array}{lcl}
\kperp^3\kpar^{-5},&\mathrm{if}&\kpar\gtrsim\kperp^{4/3},\\
\kpar^{0}\kperp^{-11/3},&\mathrm{if}&\kpar\lesssim\kperp^{4/3}.
\end{array}
\rt.
\eeq 
\Figref{fig:Em} shows these and illustrates their relationship 
to other spectra derived above. 

The 1D spectra that follow from \eqref{eq:Eminus} are
\begin{align}
E_m^{\perp-}(\kperp) &= \int\rmd\kpar\,E_m^-(\kpar,\kperp)
\sim \frac{\kperp^{-7/3}}{m^{5/2}},\\
E_m^{\parallel-}(\kpar) &= \int\rmd\kperp\,E_m^-(\kpar,\kperp)
\sim \frac{\kpar^{-2}}{m^{5/2}}.
\end{align}
Note that these are both subdominant, in $m$, to the ``$+$"-mode spectra
\exref{eq:Eperpm} and \exref{eq:Eparm}. 

\subsection{Effect of collisions}

\subsubsection{Collisional cutoff for phase-mixing modes}

\begin{figure}
\begin{center}
\includegraphics[width=0.85\textwidth]{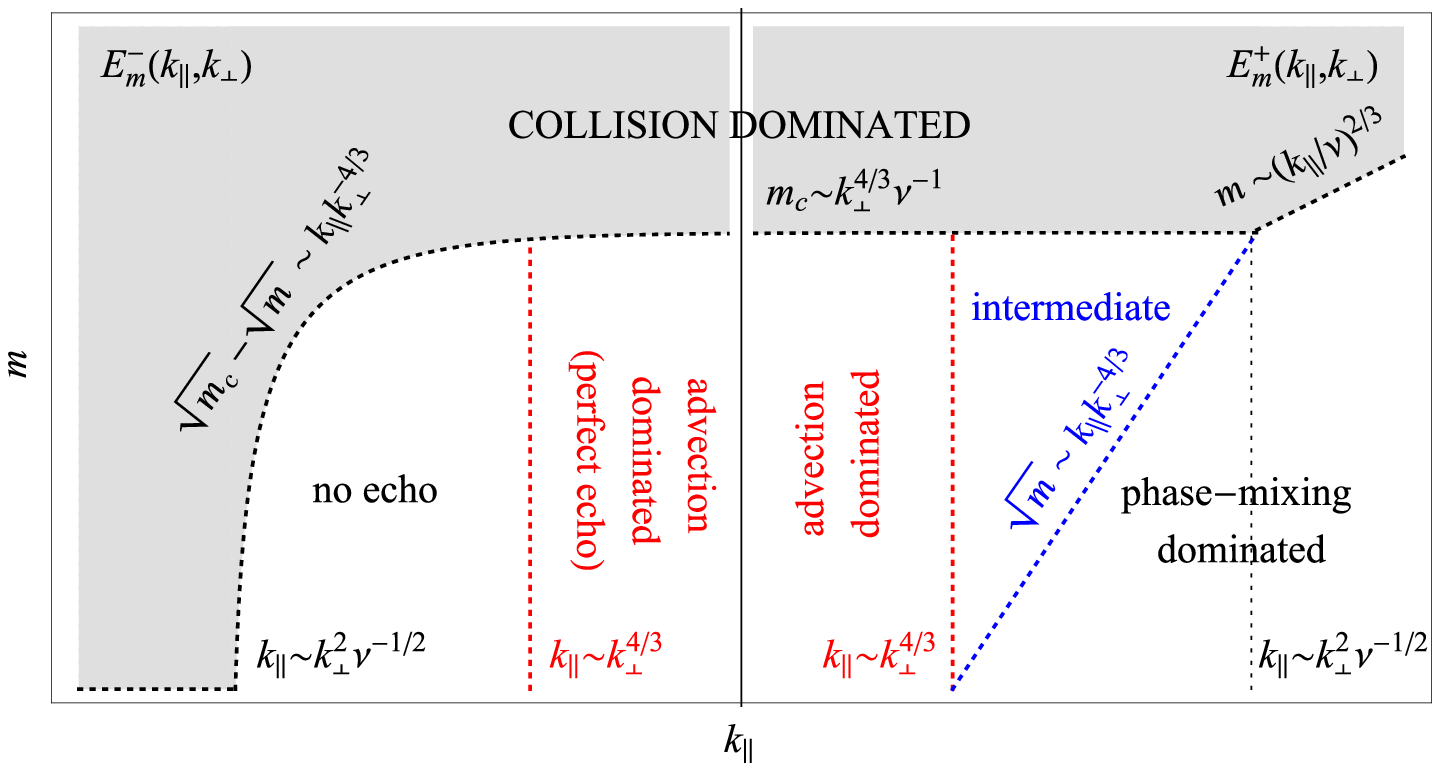}
\end{center}
\caption{Partition of phase space, viz., $(\kpar,m)$ plane at fixed $\kperp$, 
showing the collision-dominated region. Cf.\ \figref{fig:Em}(a).
All axes are logarithmic (in the left panel, $\kpar$ increases leftwards).}
\label{fig:coll_cutoffs}
\end{figure}

In the phase-mixing-dominated regime (\secref{sec:unfettered}), the collisional 
cutoff is set, in the same way as in the linear theory (\secref{sec:Hcascade}), 
by the competition between the phase-mixing rate $\sim\kpar\vth/\sqrt{m}$ and the collision 
rate $\sim\nu m$. The perturbations are collisionally damped if  
\beq
\nu m \gtrsim \frac{\kpar\vth}{\sqrt{m}}\gtrsim\kperp\uperp 
\quad\Leftrightarrow\quad
m\gtrsim \lt(\frac{\kpar}{\nu}\rt)^{2/3},\quad\kpar\gtrsim\frac{\kperp^2}{\sqrt{\nu}},
\label{eq:mc_kpar}
\eeq
giving a cutoff in Hermite space (cf.\ \eqref{eq:Cm_sln}).\footnote{In the last formula 
in \eqref{eq:mc_kpar}, we implicitly nondimensionalised the collision frequency: 
$\nu\Lpar/\vth = \Lpar/\mfp \to\nu$, so the Hermite cutoff is $\mc\sim(\kpar\mfp)^{2/3}$,
where $\mfp=\vth/\nu$ is the mean free path. 
In other words, in rescaled units, one can replace $\nu\Leftrightarrow 1/\mfp$ wherever 
this makes things more transparent.} 
In both the intermediate (\secref{sec:int}) and advection-dominated (\secref{sec:fluid_region})
regimes, the relevant comparison is between the collision rate and the nonlinear-advection rate: 
\beq
\nu m \gtrsim \kperp\uperp \gtrsim \frac{\kpar\vth}{\sqrt{m}}
\quad\Leftrightarrow\quad
m\gtrsim \frac{\kperp^{4/3}}{\nu},\quad\quad\kpar\lesssim\frac{\kperp^2}{\sqrt{\nu}}.
\label{eq:mc_kperp}
\eeq
These cutoffs are sketched in \figref{fig:coll_cutoffs}. 

At a fixed $m$, the above relations imply that there is an infrared collisional cutoff 
in the $(\kpar,\kperp)$ space: perturbations are damped if  
\beq
\kpar \lesssim \nu m^{3/2}, \quad
\kperp \lesssim (\nu m)^{3/4}.
\label{eq:cutoff_k}
\eeq
These cutoffs will not, of course, be relevant in comparison with the outer scales 
($\kparo$ and $\kperpo$; see \secref{sec:outer}) except 
at high enough $m$ or if the collision frequency approaches the characteristic
phase-mixing and nonlinear-advection rates at the outer scale. In the latter case, one 
expects some amount of free energy to drain via collisional dissipation around the outer scale, 
petering out at larger $(\kpar,\kperp)$, where the collisionless physics asserts itself
\citep[cf.][]{hatch11prl,hatch11pop,hatch13,hatch14}.   

\subsubsection{Collisional cutoff for anti-phase-mixing modes}

The collisional cutoff \exref{eq:mc_kperp} on the free-energy spectrum of 
the ``$+$" modes in the advection-dominated regime must extend to the ``$-$'' 
(anti-phase-mixing) modes because nonlinear coupling is the only source 
of the latter (see \secref{sec:fluid_region}). But the anti-phase-mixing modes 
propagate from high to low $m$ and so a zero ``boundary condition'' at high $m$ 
will be imprinted onto a region of phase space at lower $m$'s. To wit, arguing 
analogously to \secref{sec:unfettered} and considering now $\kpar<0$ in \eqref{eq:tf}, 
we note that anti-phase-mixing modes propagate along the characteristics
\beq
s = -\frac{|\kpar|\vth}{\sqrt{2}}\,t + s_0,
\label{eq:s0}
\eeq
where $s_0$ is a constant. Whatever anti-phase-mixing spectrum exists at $s=s_0$, it will 
be replicated over all $s$ satisfying \eqref{eq:s0} for times shorter than the nonlinear 
time, $t\lesssim\tauc\sim(\kperp\uperp)^{-1}\sim\kperp^{4/3}$. Assuming that $E_m^-$ 
is cut off for $s\gtrsim \scoll=\kperp^{2/3}/\sqrt{\nu}$ (\eqref{eq:mc_kperp}) and 
letting $s_0=\scoll$, we conclude that $E_m^-$ must also be cut off for 
\beq
s \gtrsim \scoll - \frac{|\kpar|\vth}{\kperp\uperp} 
\quad\Leftrightarrow\quad
\sqrt{m} \gtrsim \sqrt{\mc} - \frac{|\kpar|}{\kperp^{4/3}},
\quad \mc = \frac{\kperp^{4/3}}{\nu}.  
\label{eq:anti_cutoff}
\eeq
This implies, in particular, that there is no anti-phase-mixing energy at any $m$'s
for wave numbers satisfying 
\beq
|\kpar| \gtrsim \frac{\kperp^2}{\sqrt{\nu}}. 
\eeq
The collision-dominated region of the phase space is sketched in \figref{fig:coll_cutoffs}. 

Whereas the collisional cutoff is safely removed to infinite $m$ in the limit $\nu\to0$, 
in systems with only moderately low collision frequency, one should expect to see a
finite reduction in the anti-phase-mixing flux at higher $m$'s, 
as per \eqref{eq:anti_cutoff}.  

\subsection{Total free energy and dissipation}
\label{sec:totals}

In the linear problem, where all energy injected into the system had to be removed 
by Landau damping (meaning phase mixing followed by collisional dissipation at high $m$), 
the free energy stored in phase space in a steady state had to diverge with vanishing 
collisionality (see \eqref{eq:Wlin}) in order for the dissipation to remain 
finite (\eqref{eq:diss_lin}). In the nonlinear situation with which we are 
now preoccupied, the Hermite spectra are steep power laws and so the free energy 
will be finite and collisional 
dissipation vanish, with all of the injected energy having to be removed via dissipation 
at small spatial scales (sub-Larmor and so outside the regime of validity of this theory). 

To demonstrate this a little more quantitatively, let us repeat the calculation of the 
1D parallel spectrum (\eqref{eq:Eparm}), but now, in integrating the 2D spectrum 
over $\kperp$, we assume that there is no free energy at perpendicular 
wave numbers below the collisional cutoff $\kperp\sim(\nu m)^{3/4}$ (\eqref{eq:cutoff_k}). 
The following three cases correspond to the collisional cutoff falling into 
the phase-mixing-dominated, intermediate and advection-dominated regions, 
respectively:  
\beq
E_m^{\parallel +}(\kpar) \sim \lt\{
\begin{array}{lcl}
\displaystyle
\frac{\kpar^{-2}}{m^2} & \mathrm{if} & 
\displaystyle
m \lesssim \lt(\frac{\kpar}{\nu}\rt)^{2/3} \sim \mc,\\
&&\\
\displaystyle
\frac{\kpar^{-1}}{\nu m^{7/2}} & \mathrm{if} & 
\displaystyle
\lt(\frac{\kpar}{\nu}\rt)^{2/3}\lesssim m \lesssim \frac{\kpar}{\nu},\\
&&\\
\displaystyle
\frac{\kpar^0}{\nu^2 m^{9/2}} & \mathrm{if} & 
\displaystyle
m \gtrsim \frac{\kpar}{\nu}.
\end{array}
\rt.
\eeq
Note that the last two cases are only relevant at very high $m$
(because $\kpar\ge \kparo\sim1$, the outer scale in our units).  
Now integrating these spectra over $m$, we find that the total 
free energy in a given $\kpar$ is completely dominated by low $m$'s:
\beq
W(\kpar)\sim \int_{\sim 1}^\infty \rmd m\,E_m^{\parallel +}(\kpar) \sim \kpar^{-2}. 
\label{eq:Wnlin}
\eeq
The total collisional-dissipation rate vanishes with $\nu$:  
\beq
D(\kpar)\sim \nu\int_{\sim 1}^\infty \rmd m\,m E_m^{\parallel +}(\kpar) 
\sim \kpar^{-2}\nu\int_{\sim 1}^{\mc} \frac{\rmd m}{m}
\sim \kpar^{-2}\nu\ln \lt(\frac{\kpar}{\nu}\rt)^{2/3} \to 0
\quad\mathrm{as}\quad \nu\to 0.
\label{eq:diss_nlin}
\eeq
\Eqsand{eq:Wnlin}{eq:diss_nlin} are the nonlinear versions of \eqsand{eq:Wlin}{eq:diss_lin}, 
respectively---we see that, unlike in the linear problem, the free energy remains 
finite and collisional dissipation vanishes as $\nu\to+0$.  

\section{Conclusion}
\label{sec:concs}

\subsection{Summary of free-energy spectra and of the method of deriving~them}
\label{sec:sum}

Considering the full phase space $(\kpar,\kperp,m)$, we posited a set 
of power-law relationships for the free-energy spectra and then 
determined the scaling exponents from a combination of matching 
conditions between different regions of the phase space and 
physical arguments about the free-energy flows, constrained by 
conservation laws. The spectra are, for the phase-mixing modes 
(i.e., perturbations that propagate from low to high~$m$), 
\beq
\label{eq:sum_Eplus}
E_m^+(\kpar,\kperp) \sim \lt\{
\begin{array}{lcl}
\displaystyle
\frac{\kpar^0\kperp^{-11/3}}{m^{5/2}} & \mathrm{if} & \kpar \lesssim\kperp^{4/3}\\
&&\text{(advection dominated)},\\
\displaystyle
\frac{\kpar^{-1}\kperp^{-7/3}}{m^{5/2}} & \mathrm{if} & \kperp^{4/3} \lesssim \kpar \lesssim \sqrt{m}\,\kperp^{4/3}\\
&&\text{(intermediate)},\\
\displaystyle
\frac{\kpar^{-5}\kperp^3}{\sqrt{m}} & \mathrm{if} & \kpar \gtrsim\sqrt{m}\,\kperp^{4/3}\\
&&\text{(phase-mixing dominated)},
\end{array}
\rt.
\eeq
for the anti-phase-mixing modes (propagating from high to low~$m$),
\beq
\label{eq:sum_Eminus}
E_m^-(\kpar,\kperp) \sim \lt\{
\begin{array}{lcl}
\displaystyle
\frac{\kpar^0\kperp^{-11/3}}{m^{5/2}} & \mathrm{if} & \kpar \lesssim\kperp^{4/3}\\
&&\text{(advection dominated)},\\
\displaystyle
\frac{\kpar^{-5}\kperp^3}{m^{5/2}} & \mathrm{if} & \kpar \gtrsim\kperp^{4/3}\\
&&\text{(no echo)},
\end{array}
\rt.
\eeq
and for the ``fluid'' (low-$m$) moments, 
\beq
\label{eq:sum_Ephi}
\Eph(\kpar,\kperp) \sim \lt\{
\begin{array}{lcl}
\displaystyle
\kpar^0\kperp^{-11/3} & \mathrm{if} & \kpar \lesssim\kperp^{4/3}\\
&&\text{(advection dominated)},\\
\displaystyle
\kpar^{-5}\kperp^3 & \mathrm{if} & \kpar \gtrsim\kperp^{4/3}\\
&&\text{(phase-mixing dominated)}.
\end{array}
\rt.
\eeq
A graphical summary of these spectra is presented in \figref{fig:Em}. 

As is manifest in the above formulae, the phase space is partitioned into several 
regions, where different physics controls the distribution of the free energy.  
\begin{itemize}

\item In the {\em phase-mixing-dominated region} (\secref{sec:unfettered}), 
the phase-mixing rate is greater than the rate of nonlinear advection, 
$\kpar\vth/\sqrt{m} \gg \kperp\uperp$, 
and so whatever distribution of free energy exists at these wave numbers 
at low $m$'s will simply be propagated to larger $m$'s---this is the part 
of the wave-number space where modes are ``Landau-damped'' in the usual 
linear sense. The perpendicular spectrum in this region 
($\propto\kperp^3$; see \eqref{eq:sum_Eplus}) 
is fixed on purely kinematic grounds (\apref{ap:k3}), 
the $m$ scaling ($\propto m^{-1/2}$) is the same as in the linear 
problem, corresponding to constant Hermite flux 
(\citealt{zocco11}, \citealt{kanekar15}; see \secref{sec:Hcascade}), 
whereas the scaling exponent of the parallel spectrum ($\propto\kpar^{-a}$, $a=5$) 
is fixed by matching with the nonlinear dynamics 
(\secref{sec:int}; see \eqref{eq:rdca_final}).   

\item The continual flow of free energy into high $m$'s as described above
sets the matching condition at the {\em phase-mixing threshold}, where 
the nonlinear advection rate becomes comparable to the phase-mixing rate, 
$\kperp\uperp\sim \kpar\vth/\sqrt{m}$. The role of the nonlinear advection 
is to divert the free energy from flowing straight to higher $m$'s to 
flowing to higher $\kperp$'s. The competition between these two processes 
sets the prevailing dependence of the free energy on $m$, 
giving rise to the $m^{-5/2}$ scaling of the 2D spectra 
(\eqref{eq:sum_Eplus}; derived in \secref{sec:int_spectra} and 
\apref{ap:threshold}) and the $m^{-2}$ overall scaling 
(\eqref{eq:Eparm}; in reasonable agreement with recent 
numerical studies by \citealt{hatch13,hatch14}). The situation at 
the phase-mixing threshold is so crucial because 
the free-energy spectra rise as $\kperp$ increases and $\kpar$ decreases 
from the phase-mixing-dominated region towards the phase-mixing threshold 
and then fall beyond it, at higher $\kperp$ and lower $\kpar$, so 
it is along the phase-mixing threshold that 
the energy-containing scales in phase space lie.  
 
\item The {\em intermediate region} comprises the wave numbers 
at which the nonlinear-advection rate 
is already dominant compared to the phase-mixing rate 
of the high-$m$ moments of the distribution 
function, but not of the low-$m$ moments and, in particular, of 
the zeroth moment, $\ephi$, which is what sets the $\vE\times\vB$ flow 
velocity $\vuperp$ that is doing the advection: 
$\kpar\vth/\sqrt{m}\ll \kperp\uperp\ll \kpar\vth$. 
The energy-containing wave numbers for the flow lie along the 
{\em critical-balance} curve $\kperp\uperp\sim\kpar\vth$ 
(\secref{sec:scaling_phi})---and so  
the nonlinear interactions in the intermediate region are nonlocal 
in $\kpar$, with short-parallel-scale perturbations of the distribution 
function advected by a longer-scale flow, i.e., an effectively 2D 
velocity field (see \apref{ap:nonlocal}). A constant-flux argument 
for the free-energy cascade in $\kperp$ then fixes the 
$\kperp^{-7/3}$ scaling of the 2D free-energy spectrum in this region, 
while its $\kpar^{-1}$ scaling follows from matching to the spectra 
at the phase-mixing threshold and at the critical-balance curve 
(the second scaling in \eqref{eq:sum_Eplus}; see \secref{sec:int}).

\item Beyond the critical-balance curve, $\kperp\uperp\gg\kpar\vth$, 
the nonlinear advection is completely dominant over phase mixing, giving rise 
to the {\em advection-dominated region}. The advecting flow is now 3D and 
another constant-flux argument gives the $\kperp^{-11/3}$ scaling 
of the free-energy spectrum, whereas its $\kpar^0$ scaling is a white-noise 
spectrum deduced via a simple causality argument implying that 
perturbations with a certain perpendicular scale 
are decorrelated at parallel distances long enough that information 
cannot traverse them at the speed $\sim\vth$ over one cascade time 
corresponding to that perpendicular scale 
(the first scaling in \eqref{eq:sum_Eplus}; see \secref{sec:fluid_region}). 

\end{itemize}

The above arguments have all focused on the free energy contained in the 
perturbations that propagate from low to high $m$, i.e., ones prone 
to phase mixing (whether it is fast or slow compared to nonlinear advection). 
In a nonlinear system, an advecting flow that has a parallel spatial 
dependence, i.e., $\kpar\neq0$, can couple these perturbations to 
others that have parallel wave numbers of opposite sign and so 
will propagate from high to low $m$, a phenomenon known as 
plasma echo (\secref{sec:nlin}). Separating all perturbations 
into these ``$+$'' and ``$-$'' components (\secref{sec:unmix})
allows us to express the free energy as the sum of their spectra and its flux 
in Hermite space as proportional to the difference between these spectra 
(\secref{sec:energetics}). In the advection-dominated region, 
vigorous nonlinear coupling implies that the ``$+$'' and ``$-$'' 
spectra are the same and so, statistically, there is no free-energy flux 
between different $m$'s---i.e., the phase-mixing and the 
anti-phase-mixing energy fluxes cancel each other (see 
\secref{sec:fluid_region} and the first scaling in \eqref{eq:sum_Eminus}). 
In contrast, there is no echo effect and, therefore, no significant 
``$-$'' energy either in the intermediate region (because the flow velocity 
there is effectively 2D and so cannot couple different $\kpar$'s) 
or in the phase-mixing-dominated region (because anti-phase-mixing modes 
do not propagate to higher $m$'s). The ``$-$'' spectrum outside the 
advection-dominated region (the second scaling in \eqref{eq:sum_Eminus}, 
derived in \secref{sec:minus}) is, therefore, determined by the kinematic 
constraint giving the $\kperp^3$ scaling at long wavelengths and by the 
matching conditions along the boundary of that region---the critical-balance 
curve. 

Finally, the spectra \exref{eq:sum_Ephi} of the low-$m$, ``fluid'' moments 
are basically a continuation of the high-$m$ spectra \exref{eq:sum_Eplus} and 
\exref{eq:sum_Eminus} down to low $m$'s. Physically, since the Hermite flux 
between different $m$'s is on average shut down in the advection-dominated region, 
these scalings can be determined by assuming constant flux of the ``fluid'' part of the 
free-energy, i.e., effectively, by pretending that the turbulence is fluid-like
(\citealt{barnes11}; see \secsand{sec:in_range}{sec:scaling_phi}). Such a shortcut 
has always been tempting \citep[e.g.,][]{weiland92}, 
but was not {\em a priori} justified for a kinetic system
(\secref{sec:inconsistent}).    

\subsection{Free-energy flows}
\label{sec:flow_path}

Although this was implicit in our discussion of the partition of phase space
(\secref{sec:sum}), it is worth spelling out what path the free energy takes through it. 
Let us start from some $(m,\kpar,\kperp)$ in the {\em phase-mixing-dominated region} 
(lower right corner in the right panel of \figref{fig:Em}(a)). 
At first, the free energy will move (phase-mix) from there to higher $m$ 
(vertically towards the blue line in \figref{fig:Em}(a)) until it reaches 
the phase-mixing threshold $m\sim\kpar^2/\kperp^{8/3}$ 
(\eqref{eq:CBm} with $r=4/3$; the blue line in \figsref{fig:Em}(a,b)). 
There it enters the {\em intermediate region}, where it is advected 
by an effectively 2D velocity field (see \secref{sec:int_spectra} and \apref{ap:nonlocal}) 
to higher $\kperp$ while staying at fixed $\kpar$ 
(in the top panel of \figref{fig:Em}(b), horizontally from the blue towards 
the red line)
until it reaches the critical-balance threshold $\kperp\sim\kpar^{3/4}$
(\eqref{eq:fluid_region}; the red line in \figref{fig:Em}(b)). 
At that point it enters the {\em advection-dominated region}, 
where the advection is 3D and the energy flows along the 
critical-balance curve (diagonally upwards along the red line 
in the top panel of \figref{fig:Em}(b); see \secref{sec:fluid_region}). 
Since a 3D velocity is effective at coupling positive and negative $\kpar$'s, 
this flow of energy involves both ``$+$'' and ``$-$'' modes 
(the latter shown in the bottom panel of \figref{fig:Em}(b), 
where the critically balanced energy flow is also along the red line). 
There is not much flow of the ``$-$'' energy beyond the critical-balance 
threshold (to the left of the red line in the bottom panel of 
\figref{fig:Em}(b), or, equivalently, to the lower left of the red line 
in the left panel of \figref{fig:Em}(a)) because it nonlinearly 
couples back to ``$+$'' modes faster than it can anti-phase-mix 
to lower $m$'s.\footnote{It is possible for mode-coupling in $\kpar$ 
to combine with anti-phase-mixing to push some ``$-$'' energy towards 
larger $\kpar$, but that process is diffusive in $\kpar$ and will be 
slower than direct nonlinear coupling back into ``$+$'' modes. 
It becomes important when the advecting flow is scale-separated 
from the distribution function that is advected by it \citep{scalar}.} 

\subsection{Implications and outlook}

The free-energy distribution in phase space summarised above has several 
important properties and implications. 

The free-energy flux out of the ``fluid'' moments 
is heavily suppressed in the wave-number region bounded 
by the critical-balance curve, $\kpar\lesssim\kperp^{4/3}$, which is 
also the region that contains most of the free energy flowing through 
the inertial range. Thus, at the energetically 
relevant wave numbers of the inertial range, Landau damping is effectively absent.
The resulting Hermite spectra have steep power laws ($\propto m^{-2}$ for the 
total energy; see \eqref{eq:Eparm}) and so the total free energy contained in 
the phase space is finite, dominated by low $m$'s (the energy in the ``fluid'' 
moments) and does not diverge at vanishing collisionality (\eqref{eq:Wnlin})---in 
sharp contrast to its behaviour in the linear problem (see \eqref{eq:Wlin}). 
Furthermore, the total collisional dissipation vanishes in the 
nonlinear problem (\eqref{eq:diss_nlin}), 
again in contrast to the linear case, where the dissipation rate is finite 
and absorbs all of the energy that is injected into the system 
(\citealt{kanekar15}; see \eqref{eq:diss_lin}). 
This means that most of the dissipation occurs at small {\em spatial} scales
(i.e., beyond the Larmor scale, a region that we have left outside our detailed focus). 
This is indeed what was recently found numerically by \citet{hatch13,hatch14}: decreasing 
share of the collisional dissipation with decreasing collisionality.
Note that  finite collisionality imposes a cutoff on the free-energy 
spectra at high enough $m$'s, or, equivalently, at low enough $\kpar$ and $\kperp$ 
(\eqref{eq:cutoff_k}); 
when the collision frequency approaches the rates of phase-mixing and 
nonlinear-advection rates, a certain amount of collisional dissipation will 
occur at low wave numbers \citep[cf.][]{watanabe06,hatch11prl,hatch11pop}. 

It is inevitable that one must ask about the 
implications our results might have for the Landau-fluid closures as a viable modelling 
technique---a subject that has long been discussed and refined in the context 
of fusion plasmas 
\citep{hammett90,hammett92,hammett93,weiland92,mattor92,hedrick92,dorland93,beer96,snyder97,snyder01,snyder01gf,ramos05} 
as well as, more recently, space and astrophysical ones 
\citep{passot04,passot06,passot07,goswami05,passot12}. 
While the basic idea of the Landau-fluid approach is to include 
into fluid equations damping terms ($\sim|\kpar|\vth$) fit 
to capture correctly the linear Landau damping, 
it has long been known in this field that quantitatively these models 
work better when more Hermite moments are retained and 
this inclusion happens at the level of the highest of them \citep{smith97}. 
Considering that the free energy scales steeply with $m$, as shown 
above, it stands to reason that, at low collisionality, Landau-fluid closures that retain 
a certain finite (independent of the collision frequency) number of moments may be 
sufficient for a full characterisation of kinetic turbulence---in that already 
just this finite number of moments will be enough to capture most of the 
echo flux from phase space back to ``fluid'' moments. The Landau closure 
terms affecting the highest of the retained moments will then serve 
to regularise the problem in the energetically subdominant part 
of the wave-number space---the phase-mixing region---where the free energy 
has a shallow scaling $\sim m^{-1/2}$ (\secref{sec:unfettered}). 

There is clearly space for further development 
of this line of reasoning, leading to more quantitative prescriptions 
for capturing the echo effect within the Landau-fluid framework. 
If one thinks of these closures in the same modelling spirit as one 
does about Large-Eddy-Simulation techniques in fluid dynamics \citep{smagorinsky63}---and, 
more recently, in gyrokinetics \citep{morel11,morel12,banon14},---the $m^{-5/2}$ spectrum 
we have derived can serve the useful role of providing the signature of a well-developed 
nonlinear phase-space ``cascade,'' which, once formed, can be promptly and safely cut off by 
model dissipation terms. 

Ranging somewhat further afield, we note that spacecraft measurements 
of compressive (density and magnetic-field strength) fluctuations in the 
inertial range\footnote{These can be shown to be drift-kinetic fields 
passively advected by the turbulent velocity field $\vuperp$ associated with 
Alfv\'enic perturbations. They satisfy equations that are quite similar 
to \eqref{eq:g}, although with an additional complication that particles stream
along magnetic field that is also perturbed by the Alfv\'enic turbulence and 
so linear and nonlinear mixing are somewhat intertwined \citep{sch07,tome,kunz15}.} 
of the solar-wind turbulence 
\citep{celnikier83,celnikier87,marsch90,bershadskii04,hnat05,kellogg05,chen11,chen14}
show healthy Kolmogorov-like power-law spectra---in what is generally a $\beta\sim1$ 
plasma, where the Landau damping of such fluctuations \citep{barnes66} ought to be 
of the same order as their nonlinear cascade rates \citep{tome}. 
Similarly robust power-law spectra at sub-ion-Larmor scales 
have also been measured \citep{sahraoui09,sahraoui10,sahraoui13,alexandrova09,alexandrova12,chen10,chen13}
and found in kinetic simulations \citep{howes11prl,chang11}, even though 
Landau damping of kinetic Alfv\'en waves \citep{howes06,gary08} 
should be quantitatively noticeable at these scales \citep{howes08jgr,podesta10}. 
Whereas attempts have been made to argue that in some of these 
situations the linear damping might be weak \citep{lithwick01,howes08jgr,tome}, 
it is a tempting---and more interesting---thought that the general 
mechanism for (statistical) suppression of phase mixing in a turbulent system 
proposed here is responsible for making collisionless  
plasma turbulence in the solar wind behave in a seemingly more ``fluid-like'' fashion 
than theoreticians might have thought it had a right to do.\footnote{An immediate physically 
interesting conclusion from 
such an outcome, apart from power-law compressive spectra being theoretically legitimised, 
would be that one should not expect any ion heating associated with the 
inertial-range turbulence (see \eqref{eq:diss_nlin}), with the thermal 
fate of all turbulent energy determined at the ion Larmor scale, where
the 4D drift-kinetic phase-space cascade morphs into a more complicated 
5D gyrokinetic one \citep{tome,howes11prl,told15}.} 
A numerical and theoretical investigation of this possibility is a subject 
of our current efforts.

To conclude, the considerations presented above appear to point to a number 
of promising directions for numerical experiment and further thought. 
We hope to explore some of those in the not so distant future.\footnote{As this 
paper is going into press, the 
first dedicated numerical tests of our theory have been undertaken 
by \citet{kanekar15phd}, \citet{parker16dphil} and \citet{parker16}, 
so far broadly supporting our conclusions.} 

\begin{acknowledgments} 

We are grateful to I.~Abel, M.~Barnes, S.~Cowley, A.~Kanekar, N.~Loureiro,  
F.~Parra, C.~Staines, and L.~Stipani for many important discussions on 
this and related topics. I.~Abel, N.~Loureiro and L.~Stipani have read 
this paper in manuscript and made useful comments. Constructive critique 
from two anonymous but diligent referees have helped improve our exposition, 
for which we are thankful. 
A.A.S.\ is indebted to R.~Jeffrey for his collaboration on an unpublished early 
precursor to this project. 
J.T.P.\ was supported by the UK Engineering and Physical Sciences 
Research Council through a Doctoral Training Grant award. 
E.G.H.'s work has been carried out within the framework of the EUROfusion Consortium 
and was supported by a EUROfusion Fusion Researcher Fellowship [WP14-FRF-CCFE/Highcock]. 
The views and opinions expressed herein do not necessarily reflect those of 
the European Commission.
W.D.\ was supported by the US DoE grants DE-FG02-93ER54197 and DE-FC02-08ER54964. 
All authors are grateful to the Wolfgang Pauli Institute, University of Vienna, 
for its hospitality on several occasions. 
\end{acknowledgments} 

\appendix

\section{Long-wavelength scaling of spectra}
\label{ap:k3}

Here we review the standard argument that the spectrum of a 2D-isotropic homogeneous 
field $\ephi(\kpar,\vrperp)$ has the low-wavelength asymptotic form
\beq
\Eph(\kpar,\kperp) = 2\pi\kperp\la|\ephi(\kpar,\vkperp)|^2\ra \propto \kperp^3
\quad\mathrm{as}\quad \kperp\to0. 
\label{eq:k3}
\eeq

Setting the perpendicular-Fourier-transform conventions to be
\begin{align}
\ephi(\kpar,\vrperp) & 
= \lt(\frac{\Lperp}{2\pi}\rt)^2\int\rmd^2\vkperp e^{i\vkperp\cdot\vrperp}\ephi(\kpar,\vkperp),\\ 
\ephi(\kpar,\vkperp) &= \int\frac{\rmd^2\vrperp}{\Lperp^2}\,e^{-i\vkperp\cdot\vrperp}\ephi(\kpar,\vrperp), 
\end{align}
where $\Lperp$ is the box size, we find
\begin{align}
\nonumber
\la|\ephi(\kpar,\vkperp)|^2\ra &= 
\int\frac{\rmd^2\vr_{\perp1}}{\Lperp^2}\int\frac{\rmd^2\vr_{\perp2}}{\Lperp^2}\,
e^{-i\vkperp\cdot(\vr_{\perp1}-\vr_{\perp2})}
\la\ephi(\kpar,\vr_{\perp1})\ephi^*(\kpar,\vr_{\perp2})\ra\\
&=\int\frac{\rmd^2\vrperp}{\Lperp^2}\,e^{-i\vkperp\cdot\vrperp} C(\kpar,\rperp)
=\frac{2\pi}{\Lperp^2}\int_0^\infty\rmd\rperp\rperp J_0(\kperp\rperp)C(\kpar,\rperp),
\label{eq:phiksq}
\end{align}
where $\vrperp=\vr_{\perp1}-\vr_{\perp2}$, 
$C(\kpar,\rperp)$ is the two-point correlation function 
of $\ephi$ (which only depends on $\rperp=|\vrperp|$ because the field is statistically 
homogeneous and isotropic), and $J_0$ is the Bessel function of order zero. 

If the correlation function $C(\kpar,\rperp)$ decays sufficiently quickly with $\rperp$, 
it will restrict the integral in \eqref{eq:phiksq} to values of $\rperp$ that are smaller than 
or comparable to the perpendicular correlation length $\rperpc$ of the field $\ephi(\kpar,\vrperp)$.  
Note that $\rperpc$ will be a function of $\kpar$, so it is not necessarily the outer 
scale---in \secref{sec:scaling_phi}, we argue that it is the critical-balance 
scale, $\rperpc\sim\kperpc^{-1}\sim\kpar^{-1/r}$. 
If we now consider $\kperp\rperpc\ll1$, we may expand the Bessel function
$J_0(\kperp\rperp) = 1 - \kperp^2\rperp^2/4 + \dots$ in \eqref{eq:phiksq},
which then gives us
\begin{align}
\nonumber
\Eph(\kpar,\kperp) = 2\pi\kperp\la|\ephi(\kpar,\vkperp)|^2\ra
& = \kperp\lt[\lt(\frac{2\pi}{\Lperp}\rt)^2\int_0^\infty\rmd\rperp\rperp C(\kpar,\rperp)\rt]\\ 
& + \kperp^3\lt[-\frac{1}{4}\lt(\frac{2\pi}{\Lperp}\rt)^2\int_0^\infty\rmd\rperp\rperp^3 C(\kpar,\rperp)\rt]
+ \dots
\label{eq:E_exp}
\end{align}
The first term is proportional to $\int\rmd^2\vrperp\la\ephi(\kpar,\vrperp)\ephi^*(\kpar,0)\ra$ 
and so it vanishes if we assume that 
$\int\rmd^2\vrperp\ephi(\kpar,\vrperp)=\ephi(\kpar,\kperp=0)=0$, i.e., that there are 
no purely 1D parallel modes.\footnote{In the theory of a passive scalar,
the quantity $\int_0^\infty\rmd^2\vrperp C(\kpar,\rperp)$ is known as 
the \citet{corrsin51inv} invariant---the decay laws for a passive scalar can depend 
on whether this invariant is zero or finite because that sets the long-wavelength asymptotic
behaviour of the scalar's spectrum \citep[e.g.,][]{eyink00,sch04}. 
The fact that this asymptotic behaviour is $\sim\kperp^3$ 
in our theory, will have implications for the determination of the Hermite 
spectrum; see \apref{ap:threshold}.} 
Hence we obtain the desired result \exref{eq:k3}, with the proviso 
that $C(\kpar,\rperp)$ decays faster than $1/\rperp^4$ as $\rperp\to\infty$ and so the 
integral prefactor of $\kperp^3$ in \eqref{eq:E_exp} converges. 

\section{Nonlocal interactions in the intermediate region}
\label{ap:nonlocal}

Consider the nonlinear coupling expressed by the right-hand side of \eqref{eq:tf}, 
which we now rewrite as a wave-number convolution in both parallel and perpendicular 
directions:  
\beq
\lt(\frac{\dd \tf}{\dd t}\rt)_\mathrm{nl}\!\!\!(\kpar,\vkperp) 
= -i\vkperp\cdot\!\!\sum_{\ppar,\vpperp}\vuperp(\ppar,\vpperp)\tf(\kpar-\ppar,\vkperp-\vpperp).
\label{eq:tf_nl}
\eeq
In the intermediate wave-number region between the phase-mixing threshold and 
the critical balance, 
\beq
\sqrt{m}\,\kperp\uperp\gtrsim\kpar\vth\gtrsim\kperp\uperp
\quad\Leftrightarrow\quad
\sqrt{m}\,\kperp^r\gtrsim\kpar\gtrsim\kperp^r,
\label{eq:int_region}
\eeq
the coupling in \eqref{eq:tf_nl} must be predominantly between disparate wave numbers 
(i.e., the coupling is nonlocal) because the energy-containing wave numbers for 
$\vuperp$ are $\ppar\lesssim\pperp^r$, which lie outside the region \exref{eq:int_region}. 
There are two basic possibilities: coupling that is local in $\kperp$ but nonlocal 
in $\kpar$ and coupling that is local in $\kpar$ but nonlocal in $\kperp$. 
In analysing the rates of such interactions, we will consider $\tf$ to be at the 
phase-mixing threshold, $\kpar\sim\sqrt{m}\,\kperp^r$, and $\vuperp$ in 
critical balance, $\ppar\sim\pperp^r$. 

Suppose the perpendicular coupling is local, $\pperp\sim|\vkperp-\vpperp|\sim\kperp$. 
Then 
\beq
\ppar\sim\kperp^r\sim\frac{\kpar}{\sqrt{m}}\ll\kpar,
\eeq
so the distribution function $\tf(\kpar-\ppar)\approx\tf(\kpar)$ 
is advected by an effectively two-dimensional velocity field: back in real space, 
\eqref{eq:tf_nl} becomes
\beq
\lt(\frac{\dd \tf}{\dd t}\rt)_\mathrm{nl}\!\!\!
\approx -\vuperp(z=0,\vrperp)\cdot\vdperp\tf(\kpar,\vrperp).
\label{eq:2D}
\eeq
The rate of nonlinear advection of $\tf(\kpar)$ is, as usual, 
\beq
\kperp\uperp \sim \kperp^r.
\label{eq:loc_kperp}
\eeq
Note that as there is no coupling in $\kpar$, there can be no echo. 

Now suppose instead that it is the parallel coupling that is local, 
$\ppar\sim|\kpar-\ppar|\sim\kpar$. Then 
\beq
\pperp \sim \kpar^{1/r} \sim m^{1/2r}\kperp \gg \kperp, 
\eeq
so the distribution function $\tf$ is advected by a much-smaller-scale 
(in the perpendicular direction) velocity field.  
The net effect of such an advection will be turbulent diffusion 
of $\tf$ with the effective mixing length $\sim1/\pperp$ and 
the effective diffusion coefficient 
\beq
\DT\sim\frac{\uperp}{\pperp}\sim \pperp^{r-2} \sim \kperp^{r-2} m^{(r-2)/2r}.
\eeq 
The rate of nonlinear advection associated with this process is then
\beq
\DT\kperp^2 \sim \frac{\kperp^r}{m^{(2-r)/2r}} \ll \kperp^r, 
\eeq
provided $r<2$ (which it is, considering it will turn out to be $r=4/3$). 
This is much smaller than the local-in-$\kperp$, nonlocal-in-$\kpar$ 
advection rate \exref{eq:loc_kperp}. Thus, the latter type of interactions 
will be the dominant ones---the claim we make in \secref{sec:int_spectra}, 
which this appendix is meant to back up. 

Note that other kinds of interaction---of various degree of non-locality in both 
$\kpar$ and $\kperp$---cannot prove faster because non-locality in $\kperp$ will 
always slow down coupling (diffusion is slower than advection) while more or less 
non-locality in $\kpar$ simply makes the velocity $\vuperp$ more or less two-dimensional
compared to $\tf$, without changing the rate of advection. 

\section{Spectra near phase-mixing threshold and the free-energy decay in Hermite space}
\label{ap:threshold}

In a statistical steady state, the free-energy spectrum is independent of time 
and so described by \eqref{eq:F}: 
\beq
\frac{\kpar\vth}{\sqrt{2}}\frac{\dd F}{\dd s}
= -2\Re \sum_{\ppar} \lt\la \tf^*(\kpar)\vuperp(\ppar)\cdot\vdperp\tf(\kpar-\ppar)\rt\ra
- 2\nu s^2 F.
\eeq
Let us consider the wave numbers around the phase-mixing threshold, 
for which $\kpar\vth\sim\kperp\uperp$, so 
the phase-mixing term is comparable to the nonlinear term. 
Ignoring collisions, assuming locality in $\kperp$ (see \apref{ap:nonlocal}) 
and expanding in $\ppar\sim\kperp\uperp/\vth\ll\sqrt{m}\,\kperp\uperp/\vth\sim\kpar$, 
we have (cf.\ \eqref{eq:2D})
\beq
\frac{\kpar\vth}{\sqrt{2}}\frac{\dd}{\dd s}\lt\la|\tf(\kpar)|^2\rt\ra
\approx -\lt\la \vuperp(z=0,\vrperp)\cdot\vdperp|\tf(\kpar)|^2\rt\ra.
\label{eq:F2D}
\eeq
Formally, this looks like an equation for the spectrum of a passive 2D field $\tf(\kpar,\vrperp)$, 
parametrised by $\kpar$, advected by a 2D velocity field $\vuperp(z=0,\vrperp)$
and decaying with $s$, which plays the role of time. 
While devising a specific quantitative closure for the triple correlator in the 
right-hand side of \eqref{eq:F2D} is outside the scope of this paper, 
it is plausible that the solutions around the phase-mixing threshold 
$\kpar\vth\sim s\,\kperp\uperp$ will satisfy, roughly, 
\beq
\frac{\dd\tf^2}{\dd s} \sim \frac{\kperp\uperp}{\kpar\vth}\,\tf^2 \sim 
\frac{\tf^2}{s}
\quad\Rightarrow\quad
\tf^2(s,\kpar)\propto \frac{1}{s^\mu}.
\label{eq:fsq_mu}
\eeq
Thus, the decay must be a power law, as we indeed assumed in \eqref{eq:E_adsr}. 
This decay law is set at the ``outer scale'', which is the phase-mixing 
threshold: $\kperp\sim(\kpar/s)^{1/r}$ ($\kpar$ is fixed). Below this scale, 
i.e., at $\kperp \gg (\kpar/s)^{1/r}$, the phase-mixing term is small and the 
$\tf(s,\kpar)$ is simply cascaded subject to the constant-flux argument 
proposed in \secref{sec:int_spectra} (i.e., the right-hand side of \eqref{eq:F2D} 
must vanish to lowest order in $1/s$). This gives a spectrum of the form 
\exref{eq:E_adsr}, inheriting its decay law $\sim1/m^{\sigma'}$ from the ``outer scale''. 
The decay law of the spectrum in the advection-dominated region $\kperp\gtrsim\kpar^{1/r}$, 
\eqref{eq:Em_s}, is then the same, $\sigma=\sigma'$, via matching at the critical-balance 
curve $\kperp\sim\kpar^{1/r}$ (see \eqref{eq:apds}). 

What is the relationship between $\mu$ and $\sigma'$ and how is this scaling 
exponent determined? \Eqref{eq:fsq_mu} effectively sets the 1D 
parallel-wavenumber spectrum, i.e., as explained above, 
the free-energy content of all wavenumbers $\kperp\gtrsim(\kpar/s)^{1/r}$: 
using \eqref{eq:E_adsr}, we get 
\beq
\frac{\tf^2(s,\kpar)}{s}\sim E^\parallel_m(\kpar) 
\sim \int_{(\kpar/\sqrt{m})^{1/r}}^\infty\rmd\kperp
E_m^+(\kpar,\kperp) \sim 
\frac{\kpar^{-a'-(d'-1)/r}}{m^{\sigma'-(d'-1)/2r}},
\label{eq:Epar1D_UV}
\eeq 
so $\mu=2\sigma'-1 - (d'-1)/r$. This decay exponent, or, equivalently, $\sigma'$, 
is deduced (along with $a'$) by matching the decay law \exref{eq:fsq_mu} 
with the decay law of the total variance of $\tf^2(\kpar)$ 
contained at long wavelengths $\kperp\lesssim(\kpar/s)^{1/r}$: using 
the asymptotic form~\exref{eq:E3ar}, we get 
\beq
\frac{\tf^2(\kpar)}{s} \sim E^\parallel_m(\kpar) 
\sim \int_{0}^{(\kpar/\sqrt{m})^{1/r}}\rmd\kperp E_m^+(\kpar,\kperp)
\sim\frac{\kpar^{-a+4/r}}{m^{1/2 + 2/r}}, 
\label{eq:Epar1D_IR}
\eeq 
so $\mu = 4/r$. Matching \eqsand{eq:Epar1D_UV}{eq:Epar1D_IR}, we get 
two relations constraining $\sigma'$ and $a'$, which, combined with 
matching conditions at the critical-balance curve $\kperp\sim\kpar^{1/r}$, 
are the same as \eqsref{eq:apds}. Note that using 
the set of exponents \exref{eq:rdca_final} and \exref{eq:das_final}, we happily 
recover the 1D parallel spectrum \exref{eq:Eparm} from either of 
\eqsand{eq:Epar1D_UV}{eq:Epar1D_IR}. We also find 
that $\mu = 3$.

Note that deducing the decay law of a turbulent field by fixing the long-wavelength 
asymptotic behaviour 
of its spectrum ($E_m\propto \kperp^3$ in our case) is a standard trick of the trade 
in turbulence theory \citep[e.g.,][]{K41decay,corrsin51inv,saffman67,eyink00,sch04,davidson10,davidson13}. 

\bibliography{ssskdh_JPP}{}
\bibliographystyle{jpp}

\end{document}